\documentclass[12pt,table, svgnames, dvipsnames,doublespace]{article}
\usepackage[utf8]{inputenc}
\usepackage[top=1in,bottom=1in,left=1in,right=1in]{geometry}
\usepackage[utf8]{inputenc}
\usepackage[sort]{natbib}
\usepackage{amsmath}
\usepackage{amsfonts} 
\usepackage{tabu}
\usepackage{graphicx}
\usepackage{subcaption}
\usepackage[table, svgnames, dvipsnames]{xcolor}
\usepackage{makecell, cellspace, caption}
\usepackage{setspace}
\usepackage{fancyhdr}

\usepackage[top=1in,bottom=1in,left=1in,right=1in]{geometry}

\begin{document}

\thispagestyle{empty} \baselineskip=28pt

\begin{center}

{\LARGE{\bf Flexible cost-penalized Bayesian model selection: developing inclusion paths with an application to diagnosis of heart disease}}
\end{center}
%


\baselineskip=12pt

\vskip 2mm

\begin{center}
Erica M. Porter\footnote{\baselineskip=10pt School of Mathematical and Statistical Sciences, Clemson University, Clemson, SC, 29643, U.S.A.}, Christopher T. Franck\footnote{Department of Statistics, Virginia Tech, Blacksburg, VA, 24061, U.S.A.}, Stephen Adams\footnote{National Security Institute, Virginia Tech, Arlington, VA, 22203, U.S.A.}
\end{center}

\vskip 2mm
\begin{center}
{\large{\bf Abstract}}
\end{center}
\baselineskip=12pt


We propose a Bayesian model selection approach that allows medical practitioners to select among predictor variables while taking their respective costs into account.  Medical procedures almost always incur costs in time and/or money.  These costs might exceed their usefulness for modeling the outcome of interest.  We develop Bayesian model selection that uses flexible model priors to penalize costly predictors \textit{a priori} and select a subset of predictors useful relative to their costs.  Our approach (i) gives the practitioner control over the magnitude of cost penalization, (ii) enables the prior to scale well with sample size, and (iii) enables the creation of our proposed inclusion path visualization, which can be used to make decisions about individual candidate predictors using both probabilistic and visual tools.  We demonstrate the effectiveness of our inclusion path approach and the importance of being able to adjust the magnitude of the prior's cost penalization through a dataset pertaining to heart disease diagnosis in patients at the Cleveland Clinic Foundation, where several candidate predictors with various costs were recorded for patients, and through simulated data.

\noindent 

\baselineskip=12pt
\par\vfill\noindent
{\bf Keywords:} Bayesian model selection, cost penalty, cost-effective.

\par\medskip\noindent

\clearpage\pagebreak\newpage \pagenumbering{arabic}
\baselineskip=24pt
\doublespacing

\section{Introduction} \label{sec:intro}

Medical studies are typically expensive to conduct, with costs measured by time, money, or required expertise.  Varying costs for predictor variables arise in settings such as medical diagnoses \citep{clevelanddata}, risk calculators \citep{Struck2020,Lloyd2019, Bang2009,Ridker2007}, and healthcare quality assessments.  When collecting or analyzing data to determine which predictors are most useful, their costs should be taken into account to accommodate available budgets.  For example, accurate medical diagnoses are crucial to ensuring that patients receive information and begin treatment promptly, if necessary.  Tests and metrics available for diagnosing medical conditions, such as heart disease, can range from relatively inexpensive background questionnaires to highly sophisticated, cutting-edge diagnostic tests.  Similarly, risk calculators such as those for chronic diseases take information like easily-obtained family medical history and time-consuming updated tests and imaging to estimate the chances of disease onset.  While costs are ubiquitous in gathering medical information and data, few statistical variable selection methods address the cost of individual predictor variables to help medical practitioners decide which to obtain.  Perhaps surprisingly, medical data are often reported without their associated costs \citep{bolon2014framework}.  Some methods exist to identify a subset of predictors with lower costs.  However, to the best of our knowledge, none of the existing methods can alone provide practitioners easy, considerable control to change the impact cost has on selection results, output readily interpretable probabilities and model parameters, and create a convenient visual to compare many different cost-adjusted analyses at once.  We propose a Bayesian model selection approach that introduces a tuning parameter to a cost-penalizing prior on predictors and produces an inclusion path for the practitioner to visually examine the predictive power of predictors relative to their costs as the cost penalization is increased or decreased.

The idea of model selection that accounts for cost has been studied in a few areas of the statistical literature.  Most important for our proposed method, when candidate predictors have different costs required to collect them, \cite*{Fouskakis2009} proposed a model prior that penalizes individual candidate predictors \textit{a priori} based on their costs, with an application to quality of healthcare assessment.  We refer to the prior developed by \cite{Fouskakis2009} as the FND prior, for the three authors of the prior.  Bayesian model selection using the FND prior leads to selection of a less costly subset of predictors when compared to Bayesian model selection with no regard to cost. However, we have found that cost penalization from the FND prior does not always scale appropriately with sample size.  Namely, as sample size increases, the cost penalization provided by the FND prior is overpowered and may not impact selection.   

\begin{figure}[h!]
    \centering
    \includegraphics[width=0.9\linewidth]{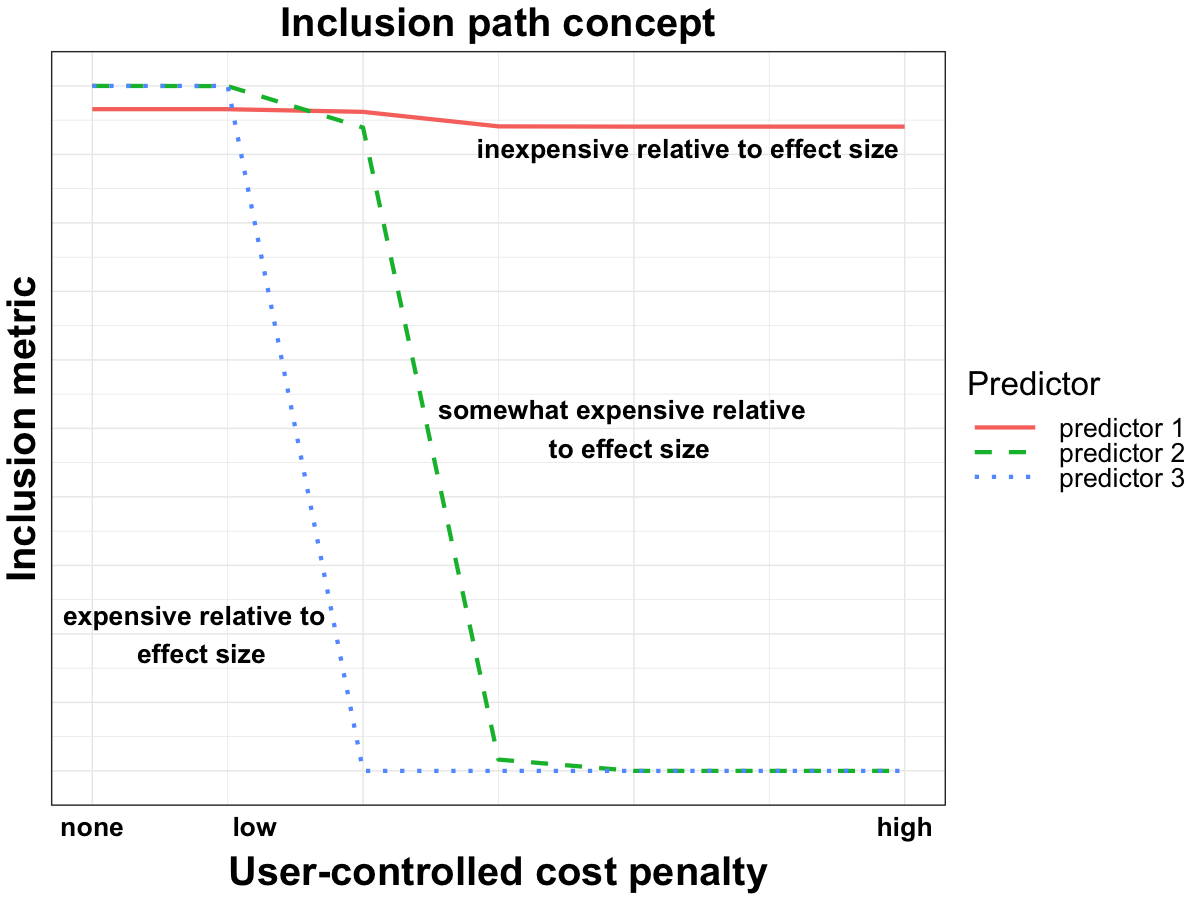}
    \caption{Diagram prefacing our proposed inclusion path.  The practitioner controls cost penalization to accommodate a budget and uses the inclusion metric value to study how each predictor's importance for modeling the outcome changes as cost is penalized differently.}
    \label{fig:schema}
\end{figure}

In Section \ref{sec:data_methods} we propose an extension of the FND prior that introduces a tuning parameter to adjust the level of cost penalization for candidate predictors, providing necessary flexibility for medical practitioners to specify the cost penalization for their problem and the decision at hand. The tuning parameter gives the practitioner the ability to directly control the amount of cost penalization and create an inclusion path that visualizes the impact of different cost penalizations on the selection of candidate predictors.  For illustration, Figure \ref{fig:schema} shows the idea: the practitioner controls the cost penalization via tuning parameter on the horizontal axis.  The y-axis indicates the value of a chosen inclusion metric for each candidate predictor at different levels of cost penalization the practitioner wishes to study.  As the practitioner increases the magnitude of cost penalization, the inclusion metric will tend to decrease for predictors whose cost is high relative to their effect size (i.e. predictor 2 and predictor 3 in Figure \ref{fig:schema}).  For our method, we choose to use posterior inclusion probabilities for each predictor as the inclusion metric.  Our method can accommodate costs recorded in terms of money, time, equipment, computations, or other measures depending on the medical application, each with the consequence of increasing the burden on overall resources.

Several methods based on decision theory have also been proposed to penalize for predictor costs.  For example, \cite{Brown1999} developed a decision-theoretic approach for multivariate linear regression which added to a quadratic loss function a terminal cost function representing the cost of keeping a particular subset of candidate predictors.  \cite{Fouskakis2008} proposed a decision-theoretic approach for binary outcome generalized linear models that appended a data collection utility component based on marginal predictor costs to the expected utility function, and they applied this utility function to several stochastic optimization algorithms.  Recently, \cite{Miyawaki2021} added a cost function to the traditional predictive loss \citep{Lindley1968} and apply Bayesian model averaging (BMA) first over purchased predictors and then by marginalizing over potential unpurchased predictors via MCMC.  They find that the latter approach performs better than standard BMA but introduces additional sensitivity in prior specification and requires further subjective prior information and assumptions, such as the joint distribution of unobserved predictors.  In another MCMC-based approach, \cite{FouskakisMCMC} used a reversible jump MCMC to search the model space constrained to models whose total predictor costs fall below a threshold.  Machine learning is another area that has seen some development of cost-penalized methods that may be adapted for medical applications \citep{elkan2001foundations, fan1999adacost, cohn1996active, settles2009active, bolon2014framework, kong2016beyond, ling2004decision, zhou2016cost, adams2016feature}.  In contrast, our method provides a single user-controlled tuning parameter to adjust the magnitude of cost penalization on candidate predictors to produce multiple cost-penalized analyses and produce probabilities for all candidate models and predictors.

The FND prior, which our proposed method extends, penalizes costly predictors relative to a minimum (baseline) cost, and \cite{Fouskakis2009} used the prior to develop a cost-adjusted selection approach which results in a generalized version of BIC.  \cite{Fouskakis2009} developed cost-adjusted BIC to select among sickness indicators for predicting death within $30$ days due to pneumonia.  \cite{Fouskakis2009} compared their selection results to those in which a uniform prior is used for all predictors and models.  The latter approach, which \cite{Fouskakis2009} call a benefit-only analysis, as it ignores costs, selects a more costly model when applied to a set of $n=2,532$ pneumonia patients.

\begin{figure}[h!]
    \centering
    \includegraphics[width=0.8\linewidth]{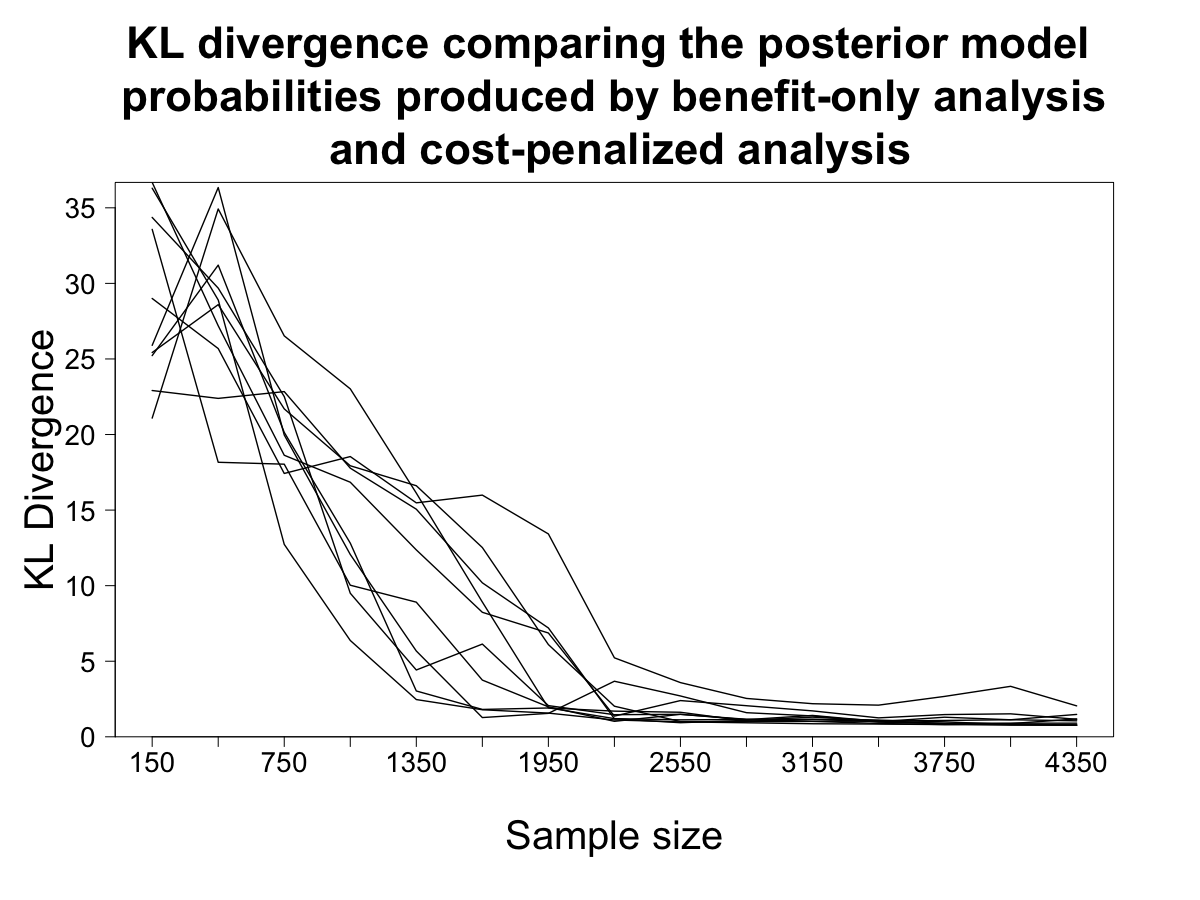}
    \caption{KL divergence calculated between the sets of posterior model probabilities produced using the FND prior and benefit-only model selection for $10$ collections of data sets where 300 observations were recursively added to 10 initial data sets of size $n=150$.  These data were generated from the linear logistic regression setting considered by \cite{Fouskakis2009}.  See Section \ref{subsec:sim_study} for more details.  The KL divergence between the posterior model probabilities produced by the two approaches decreases as the sample size grows for all $10$ collections of data sets increasing in size, with 7 of the $10$ near $0$ by $n=2500$, indicating that as sample size increases, the impact of the cost penalty on the model selection is reduced.}
    \label{fig:KL_10sets}
\end{figure}

When applying the FND prior to other data sets, we have found that the FND prior can lead to selection of less costly models when predictor costs differ, but the cost penalization is not appropriate or sufficient for all sample sizes.  In fact, at large sample sizes, the cost penalization imposed by the FND prior greatly diminishes, often causing the resulting cost-penalized model selection to closely resemble that of a standard benefit-only analysis.  To establish this phenomenon, we calculated the Kullback-Leibler (KL) divergence, which measures the difference between two probability distributions, between the sets of posterior model probabilities for all candidate models produced by cost-penalized analysis with the FND prior and the benefit-only selection approach as sample size increases for several simulated data sets.  Since we consider the case of linear logistic regression, for which there is no closed form integrated likelihood, we examine the KL divergence comparing posterior model probabilities from the two approaches, where all posterior model probabilities are obtained after applying a Laplace approximation to the integrated likelihoods.  We generated $10$ data sets of size $n=150$ with $p=9$ total candidate predictors with different costs as detailed in Section \ref{subsec:sim_study} and calculated the KL divergence between the two sets of posterior model probabilities produced by the benefit-only and FND selection approaches for each data set.  Then, to imitate collection of additional data, we recursively added $300$ observations to the initial $10$ data sets, obtaining model selection results and calculating KL divergence between the posterior model probabilities produced by the two selection approaches for each new (larger) data set.  Figure \ref{fig:KL_10sets} plots the KL divergence between the posterior model probabilities produced by the two approaches as sample size increases for the $10$ collections of data sets increasing in size.

Figure \ref{fig:KL_10sets} shows that the KL divergence value between the posterior model probabilities for all candidate models produced by the cost-penalized and benefit-only methods approaches $0$ as the sample size increases.  Thus, larger sample sizes lead cost-penalized Bayesian model selection using the FND prior to select a model with structure and cost similar to that of a benefit-only approach.  In many cases, this occurrence may not be ideal, since the user adopted the FND prior to control/reduce cost from the selected predictors.  But when the sample size of the available data is large, the cost-penalizing ability of the FND prior diminishes and the approach recommends the costly benefit-only model the user was hoping to avoid.  Ironically, collecting more observations, which often inherently increases medical study costs, can dilute or cancel out the penalization for costly predictors.  This phenomenon may lead practitioners and medical researchers to plan for studies that are more expensive than necessary.  In this paper we extend the FND prior by proposing simple functions for adjusting the cost penalization according to a tuning parameter.  These functions extend the useful FND prior and maintain the property of invariance to cost conversions, making them widely applicable.  Our inclusion path approach highlights the change in posterior inclusion probabilities for individual candidate predictors at different levels of the cost penalization according to the functions we suggest.  The inclusion path weighs the modeling ability against the cost of the predictors.  We demonstrate the utility of our adjusted cost penalization and inclusion path approach first with simulated data and then a data set collected at a medical clinic on Cleveland, Ohio, where the response is the presence of heart disease.  There are $13$ candidate predictors relevant to diagnosing heart disease available for each patient, with widely varying costs \citep{clevelanddata}.  A benefit-only approach selects many costly predictors that do not appreciably improve modeling of heart disease.  We show that our adjusted cost penalization can be used to select models with reduced costs per patient, which can help to meet hospital or insurance budgets, while still retaining cost-effective predictors for physicians to diagnose a critical condition like heart disease.  Our functions that adjust the cost penalization extend the utility of the FND prior.  The resulting inclusion path plots the changing impact of candidate predictors where the practitioner now has input over the magnitude of cost penalization.

The remainder of this paper is organized as follows.  Section \ref{sec:data_methods} introduces cost-penalized Bayesian model selection, describes the FND prior, details our functions and properties for adjusting the cost penalization on predictors, outlines our inclusion path approach, and explains the setup for our simulated data.  Section \ref{sec:results} demonstrates the utility of the FND prior and presents our adjusted cost penalization and inclusion path approach, first using simulated data and then when applied to the Cleveland heart disease data.  All selection results for our method are compared to results produced by the FND prior.  Section \ref{sec:discussion} summarizes the impact of our findings and outlines avenues for potential future research and applications in cost-penalized model selection.

\section{Data and Methods} \label{sec:data_methods}

\subsection{Bayesian model selection}
We consider Bayesian model selection with the goal of penalizing candidate predictors based on their costs through a flexible class of model priors.  Each candidate predictor has an associated cost.  We use Bayesian model selection to obtain posterior model probabilities for each possible combination of predictors, with the goal being to select a subset of predictors that accurately classify the outcome while penalizing on the basis of the costs of those predictors.  For $p$ candidate predictors, the corresponding model space is $\mathcal{M}=\{0,1\}^p= \{M_1, M_2 \ldots ,M_K\}$, where $K=|\mathcal{M}|=2^p$ is the total number of candidate models.  To compare two models, say $M_1$ and $M_2$, we may use the Bayes factor, $BF_{12}$, that is defined as the ratio of the two models' integrated likelihoods:

\begin{equation} \label{eq:standard_bf}
    BF_{12} = \frac{p(\textbf{Y}|M_1)}{p(\textbf{Y}|M_2)}.
\end{equation}



Then the posterior model probability of a single model $M_{\ell}$ in the model space can be found using Bayes' Rule:

\begin{equation} \label{eq:post_prob}
    P(M_{\ell}|\textbf{Y}) = \frac{p(\textbf{Y}|M_{\ell})P(M_{\ell})}{\sum_{k=1}^K p(\textbf{Y}|M_k)P(M_k)}= \bigg(\sum_{k=1}^K BF_{k \ell} \times \frac{P(M_k)}{P(M_{\ell})}\bigg)^{-1},
\end{equation}

\noindent where $P(M_k)$ is the prior model probability for model $M_k$. We now describe the FND prior in Section \ref{subsec:cost_mod_select}.

\subsection{Cost-penalizing model selection} \label{subsec:cost_mod_select}
\cite{Fouskakis2009} developed the FND prior for variable selection for the linear logistic regression setting, with the goal of optimizing selection of sickness indicators for improving quality of health care assessments while penalizing expensive candidate predictors.  Their motivating data set from the RAND Corporation was used to select from a list of $83$ sickness indicators to model patient deaths within $30$ days of admission due to pneumonia, where costs were measured as the time required to observe/record predictors for each patient \citep{Keeler1990}.  In particular, both the benefit-only analysis and cost-penalized analysis using the FND prior selected $13$ predictors from $83$ total candidate predictors, but the total cost of the two sets of selected predictors were 22.5 and 9.5 (in minutes), respectively, resulting in a cost reduction of more than 50\%.  Their approach is invariant to cost conversions, devised a penalty related to BIC relative to a baseline cost, and can be used to reproduce the traditional BIC when all predictor costs are equal.  Further, by using expressions based on posterior model odds, \cite{Fouskakis2009} were able to produce posterior model probabilities from a generalized cost-adjusted version of the BIC, which shares many similarities with the approximations and behaviors of the traditional BIC.

Following \cite{Fouskakis2009}, we consider the linear logistic regression setting in this paper, later applying it to the binary diagnosis of heart disease.  For convenience, we use the notation of \cite{Fouskakis2009} for indicator functions and predictor costs.  We use a linear logistic regression model where $Y_i \in \{0,1\}$ and $X_{ij}$ denotes the $j^{th}$ predictor for observation $i$, where $i=1, \ldots, n$ and $j=0, \ldots, p$.  Let $\gamma_j$ be an indicator that is equal to $1$ if predictor $X_j$ is included in the model and $0$ if it is not.  Then $\boldsymbol{\gamma}$ is a length $p$ vector of $0$'s and $1$'s indicating whether each of the $p$ predictors are in the model or not.  Note that the intercept is included in every model so $X_{i0}=1$ and $\gamma_0=1$ for all models.  Then the Bayesian modeling framework is

\begin{equation} \label{eq:logistic_reg}
\begin{split}
& (Y_i|\boldsymbol{\gamma}) \overset{\text{indep}}{\sim} \text{Bernoulli}[p_i(\boldsymbol{\gamma})],
 \\
 & \eta_i(\boldsymbol{\gamma}) = \log\Big[\frac{p_i(\boldsymbol{\gamma})}{1-p_i(\boldsymbol{\gamma})}\Big] = \sum_{j=0}^p \beta_j \gamma_j X_{ij},\\
 & \eta(\boldsymbol{\gamma}) = \boldsymbol{X}\text{diag}(\boldsymbol{\gamma})\beta = \boldsymbol{X}_{\boldsymbol{\gamma}}\boldsymbol{\beta}_{\boldsymbol{\gamma}},
\end{split}
\end{equation}

\noindent where $\boldsymbol{X}_{\boldsymbol{\gamma}}$ and $\boldsymbol{\beta}_{\boldsymbol{\gamma}}$ are the design matrix and vector of regression coefficients for the specific model with predictors in $\boldsymbol{\gamma}$.   For the vector of regression coefficients $\boldsymbol{\beta}_{\boldsymbol{\gamma}}$, we assign a Gaussian prior distribution $\pi(\boldsymbol{\beta}_{\boldsymbol{\gamma}})$ with form  $N(\boldsymbol{\mu}_{\boldsymbol{\gamma}}, \Sigma_{\boldsymbol{\gamma}})$. We assume the same prior for $\boldsymbol{\beta}_{\boldsymbol{\gamma}}$ as derived by \cite{Fouskakis2009}, so \textit{a prior} $\boldsymbol{\beta}_{\boldsymbol{\gamma}}$ is distributed as:

\begin{equation} \label{eq:beta_prior}
    \boldsymbol{\beta}_{\boldsymbol{\gamma}}| \boldsymbol{\gamma} \sim N(\boldsymbol{0}, 4n(\boldsymbol{X}_{\boldsymbol{\gamma}}^T\boldsymbol{X}_{\boldsymbol{\gamma}})^{-1}).
\end{equation}

The cost-penalized prior by \cite{Fouskakis2009} for a single $\gamma_j$ is proportional to:

\begin{equation}
    P(\gamma_j) \propto \exp\Big[-\frac{\gamma_j}{2}\Big(\frac{c_j}{c_0}-1\Big)\log n\Big],
\end{equation}

\noindent where $c_j$ is the marginal cost per observation for predictor $X_j$ and $c_0=\text{min}\{c_j, j=1,\ldots,p\}$ is defined as the baseline cost, corresponding to the cheapest (least expensive) candidate predictor.  Then the FND prior for a particular model corresponding to $\boldsymbol{\gamma}$ follows as:

\begin{equation} \label{eq:modprior}
    P(\boldsymbol{\gamma}) = \exp\bigg\{-\frac{1}{2}\log n \sum_{j=1}^p \gamma_j \bigg(\frac{c_j}{c_0}-1\bigg)-\sum_{j=1}^p \log \big[n^{-\frac{1}{2}(\frac{c_j}{c_0}-1)}+1\big]\bigg\}.
\end{equation}

A natural comparator to \eqref{eq:modprior} is a uniform prior on the model space:

\begin{equation} \label{eq:unif_priors}
    P(\boldsymbol{\gamma}) = \frac{1}{2^p},
\end{equation}

\noindent which produces the benefit-only selection when used with Equation \eqref{eq:post_prob}.  We use a Laplace approximation to approximate the integrated likelihoods with distribution \eqref{eq:logistic_reg} and prior \eqref{eq:beta_prior}, and we obtain posterior model probabilities for all models in $\mathcal{M}$ as in Equation \eqref{eq:post_prob}.  

We develop our inclusion path approach, presented in Section \ref{subsec:inclusion_path_algo}, using posterior inclusion probabilities for the candidate predictors.  The posterior inclusion probability for predictor $X_j$ is obtained by adding up the posterior model probabilities from each model containing $X_j$, i.e. where $\gamma_j=1$:

\begin{equation} \label{eq:post_inclusion_prob}
    P(\gamma_j=1|\textbf{Y}) = \sum_{\boldsymbol{\gamma}_k \in \mathcal{M}|\gamma_j=1} P(\boldsymbol{\gamma}_k|\textbf{Y}).
\end{equation}

Equation \eqref{eq:modprior} uses the marginal cost $\{c_j, j=1, \ldots p\}$ of each candidate predictor, which is the cost of collecting that predictor for a single observation.  When adhering to an overall budget, the practitioner may also want to consider the total cost per observation or the total cost of the model.  The total cost per observation is equal to the sum of the marginal cost of all predictors in the model, i.e. $\sum_{j=1}^p c_j\gamma_j$, and the total cost of the model for the given observations is $n \times \sum_{j=1}^p c_j\gamma_j$.

Note that our approach penalizes predictors based on their predetermined costs.  Therefore, the resulting posterior probabilities that we discuss are cost-adjusted probabilities rather than probabilities in the traditional Bayesian sense of the true model.  These cost-adjusted probabilities can be used to determine whether a model contains cost-effective or efficient predictors relative to other models when provided with the desired level of cost-penalization from the user, rather than being interpreted as the chance of a model being the true model.  For the remainder of this paper, we refer to any posterior probabilities obtained under a cost penalty on the predictors as `cost-adjusted posterior probabilities'.  The exception is the benefit-only model selection approach, which we show can be obtained by setting our proposed tuning parameter equal to 0, and which produces posterior probabilities that are not cost-adjusted.

\subsection{Adjusted cost-penalizing functions}\label{subsec:inclusion_path_setup}

The cost penalization from the FND prior in Equation \eqref{eq:modprior} increases with sample size, i.e.\@, through the $\log(n)$ term.  However, as demonstrated in Figure \ref{fig:KL_10sets}, the cost penalization does not always grow quickly enough with $n$, so the FND prior may not be sufficient for all data sets or budgets, particularly in the case of large sample sizes.  We propose an extension of the FND prior that gives the practitioner more flexibility when penalizing predictors based on their costs \textit{a priori}.  One way to improve the flexibility of the FND prior is the ability to adjust the cost penalization rather than relying on a fixed penalization for every application with cost.  If the cost of the model selected based on existing data exceeds a current or future budget, it would be crucial to be able to increase the cost penalization to find an effective but affordable model.  For other data, it might be the case that the model selected using the FND prior does not provide the desired performance and then it may be crucial to decrease the cost penalization to allow for selection of additional predictors that would improve the model's overall performance while still penalizing for high costs, just to a lesser extent.  To give practitioners more control over the FND prior so that it may scale to their problem, we propose functions of the cost ratio $c_j/c_0$ according to a tuning parameter $b$.  These functions change the magnitude of cost penalization in the FND prior.  

We propose functions $g(c_j/c_0,b)$ to adjust the cost ratio according to tuning parameter $b$ that satisfy the following properties:

\begin{enumerate}
    \item[(a)] $b=0$ implies $g(c_j/c_0,b)=1$ for all $j=1, \ldots, p$, reducing the prior to the uniform model prior \eqref{eq:unif_priors} and resulting in a benefit-only model selection. 
    \item[(b)] $b=1$ makes $g(c_j/c_0,b)$ equal to $c_j/c_0$ and reproduces the FND prior as seen in Equation \eqref{eq:modprior}.
    \item[(c)] When $b>1$, $g(c_j/c_0,b)$ penalizes predictors with costs $c_j>c_0$ more highly than the FND prior in \eqref{eq:modprior}. When $0 < b < 1$, $g(c_j/c_0,b)$ penalizes predictors with costs $c_j>c_0$ less than the FND prior but more than a benefit-only analysis.  Higher values of $b$ increase $g(c_j/c_0,b)$ and the resulting cost penalization, leading to lower cost-adjusted prior inclusion probabilities for predictors with costs above the baseline.
    \item[(d)] When $c_j=c_0$, the $j^{th}$ candidate predictor cost is the same as the baseline cost.  Changing $b$ does not introduce/increase penalization on any of the predictors with baseline cost.
\end{enumerate}

We propose to use a model prior of the form \eqref{eq:modprior} where the cost ratio $c_j/c_0$ is replaced with the cost ratio function $g(c_j/c_0,b)$, as follows

\begin{equation}\label{eq:adjusted_prior}
    P(\boldsymbol{\gamma}) = \exp\bigg\{-\frac{1}{2}\log n \sum_{j=1}^p \gamma_j \bigg(g\bigg(\frac{c_j}{c_0},b\bigg)-1\bigg)-\sum_{j=1}^p \log \big[n^{-\frac{1}{2}\big(g\big(\frac{c_j}{c_0},b\big)-1\big)}+1\big]\bigg\}.
\end{equation}

In general, monotone functions of the cost ratio and tuning parameter $b$ are well-suited for our cost-adjusted model prior and proposed inclusion path.  Here we study an exponential function of the cost ratio according to the tuning parameter.  Another sensible choice is a linear function of the cost ratio, which we describe in the Supplementary Material.



\color{black}
For a given value of the tuning parameter $b$, the exponential function of the cost ratio for predictor $X_j$ is

\begin{equation} \label{eq:exp_adjust_formula}
 g\bigg(\frac{c_j}{c_0},b\bigg) = \bigg(\frac{c_j}{c_0}\bigg)^b,
\end{equation}

\noindent which satisfies properties (a)-(d).   We call the prior that uses the exponential function of the cost ratio function in \eqref{eq:exp_adjust_formula} the exponential cost prior (ECP).  That is, the ECP is model prior with form \eqref{eq:adjusted_prior} where $g(c_j/c_0,b)=(c_j/c_0)^b$.

\subsection{Inclusion paths}\label{subsec:inclusion_path_algo}
Using the ECP and varying $b$, we create an inclusion path for each candidate predictor.  To do this, we compute and plot cost-adjusted posterior inclusion probabilities for each of the candidate predictors across multiple values of $b$.  
See the Supplementary Material for detailed steps for constructing the inclusion path.  The inclusion path visualizes a probabilistic way to learn the order in which the candidate predictors would be chosen or discarded, e.g.\@ according to some threshold, if a different magnitude of cost penalization is used.  The costs $\{c_j, j=1, \ldots p\}$ remain unchanged; $g(c_j/c_0,b)$ changes how heavily candidate predictors with larger costs are penalized \textit{a priori}.  Our inclusion path technique is inspired by the interpretable path diagrams like LASSO and ridge \citep{Tibshirani1996,Hoerly1970}.  From the plot of inclusion paths, the practitioner can all at once study candidate predictors' posterior inclusion probabilities for the benefit-only analysis and cost-adjusted posterior inclusion probabilities for the FND prior and a wide range of the ECP with differing $b$ values.  This is because we formulated the ECP to satisfy properties (a)-(d) so that the established uniform priors and the FND prior can be easily recreated from the ECP.


\section{Results} \label{sec:results}

\subsection{Simulation study settings} \label{subsec:sim_study}

To study the model selection behavior resulting from the FND prior and ECP, consider a simulated data set of size $n=150$ with form \eqref{eq:logistic_reg} and vector of regression coefficients

\begin{equation} \label{eq:beta_vec}
\boldsymbol{\beta}=(1, 0, 0, 0, 0.5, 0.5, 0.5, 0.8, 0.8, 0.8)^T,    
\end{equation}
 
\noindent and corresponding cost vector 
\begin{equation} \label{eq:cost_vec}
\boldsymbol{c}=(1, 3, 9, 1, 3, 9, 1, 3, 9),    
\end{equation}

\noindent representing predictors with all combinations of null, smaller, and larger effect sizes and baseline, cheap, and expensive costs.  The predictors $X_1, \ldots, X_9$ are generated independently from a $N(0,1)$ distribution.  The vector of regression coefficients \eqref{eq:beta_vec} and predictor costs \eqref{eq:cost_vec} were also used to generate the $10$ series of data sets increasing in size used to calculate KL divergence values between the the posterior model probabilities (benefit-only and cost-adjusted) resulting from the benefit-only analysis and the FND prior in Section \ref{sec:intro}.

In practice, medical costs can be measured and recorded according to money, time, labor, and many other quantities.  The FND prior and our ECP extension are both invariant to cost units/conversions since the cost penalizations are expressed relative to the baseline cost.  Often the goal may be to focus on efficiency rather than minimizing an overall cost, as this may better serve a patient or hospital operations.  Therefore, we choose to consider the costs for our simulated data as the time in minutes required to collect each predictor for one individual.  Thus, our simulation setting has $512$ candidate models with costs ranging from $0$ minutes (corresponding to the intercept-only) and $39$ minutes per observation.  Section \ref{sec:results} presents selection results for data generated as above using Bayesian model selection as described in Section \ref{sec:data_methods} with uniform priors, the FND prior, and the ECP applied to the model space.

\subsection{Simulation results using FND prior}
Consider a data set of size $n=150$ generated as described in Section \ref{subsec:sim_study}.  Table \ref{table:n150_select} highlights the difference between the selection results from a benefit-only approach and selection using the FND prior.  For the following sections, we will describe both MAP model, i.e. the model selected using the maximum posterior model probability (benefit-only or cost-adjusted) and the median probability model, which consists of predictors' whose posterior inclusion probability (benefit-only or cost-adjusted) is at least 0.5 \citep{Barbieri2004}.  For the data sets described in Tables \ref{table:n150_select} and \ref{table:n600_select}, the MAP model and the median probability model are the same.  We can see from Table \ref{table:n150_select} that the benefit-only model selection approach selects the model containing $X_5$ through $X_9$, i.e. the cheap and expensive predictors with smaller effect size and all three of the predictors that have the larger effect size.  This model has a total cost of 25 minutes per observation.  The posterior model probability for this benefit-only model is 0.332, moving off of the prior probability of 0.002, and the corresponding C-statistic, defined as the area under the receiver operating characteristic (ROC) curve, is 0.84.  Meanwhile, model selection using the FND prior on the model space selects the model with the following predictors: $X_7$ (baseline predictor with larger effect size) and $X_8$ (cheap predictor with larger effect size), for a total cost of 4 minutes per observation, a $84\%$ reduction in cost from the benefit-only model.  The cost-adjusted posterior model probability for this model is 0.491, moving off of its prior probability of 0.0008, and the corresponding C-statistic is 0.782.  We can see that the FND prior leads to a remarkable reduction in the cost per observation for its selected model and successfully prioritizes predictors with larger effect sizes while it considers their costs, while incurring modest loss in accuracy as measured by the C-statistic.

\begin{center}
\begin{table}[h!]
\centering
\begin{tabular}{ cc }   
Benefit-only selection & FND selection \\  
$n=150$  & $n=150$ \\  
\begin{tabular}{ |c|c|c|c| } 
 \hline
 \textbf{Predictor} & \textbf{Effect} & \textbf{Cost}& \textbf{Cost} \\ 
 & \textbf{size} & \textbf{level} & \textbf{(mins.)}\\
 \hline
 $X_1$ & null & baseline & 1 \\ 
 \hline
 $X_2$ & null & cheap & 3 \\
 \hline
 $X_3$ & null & expensive & 9 \\
 \hline
 $X_4$ & smaller & baseline & 1 \\ 
 \hline
 \rowcolor{Dandelion}
 $X_5$ & smaller & cheap & 3 \\   
 \hline
 \rowcolor{Dandelion}
 $X_6$ & smaller & expensive & 9 \\ 
 \hline
 \rowcolor{Dandelion}
 $X_7$ & larger & baseline & 1 \\ 
 \hline
 \rowcolor{Dandelion}
 $X_8$ & larger & cheap & 3 \\   
 \hline
 \rowcolor{Dandelion}
 $X_9$ & larger & expensive & 9 \\ 
 \hline
 \rowcolor{white}
 \multicolumn{3}{c}{} & \multicolumn{1}{c}{\underline{\textbf{25/obs}}} \\
\end{tabular} &  
\begin{tabular}{ |c|c|c|c| } 
 \hline
\textbf{Predictor} & \textbf{Effect} & \textbf{Cost}& \textbf{Cost} \\ 
 & \textbf{size} & \textbf{level} & \textbf{(mins.)}\\ \hline
 $X_1$ & null & baseline & 1 \\ 
 \hline
 $X_2$ & null & cheap & 3 \\
 \hline
 $X_3$ & null & expensive & 9 \\
 \hline
 \rowcolor{white}
 $X_4$ & smaller & baseline & 1 \\ 
 \hline
 $X_5$ & smaller & cheap & 3 \\   
 \hline
 \rowcolor{white}
 $X_6$ & smaller & expensive & 9 \\ 
 \hline
 \rowcolor{Dandelion}
 $X_7$ & larger & baseline & 1 \\ 
 \hline
 \rowcolor{Dandelion}
 $X_8$ & larger & cheap & 3 \\   
 \hline
 \rowcolor{white}
 $X_9$ & larger & expensive & 9 \\ 
 \hline
 \rowcolor{white}
 \multicolumn{3}{c}{} & \multicolumn{1}{c}{\underline{\textbf{4/obs}}}\\
\end{tabular} \\
\end{tabular}
\caption{Candidate predictors and their corresponding effect sizes and costs. Rows colored in orange  \fcolorbox{black}{Dandelion}{\rule{0pt}{5pt}\rule{5pt}{0pt}}\quad are the predictors selected by (left) the benefit-only analysis and (right) selection using the FND prior for a data set of size $n=150$.  At this small sample size, the benefit-only model costs more than $4$ times the model selected using the FND prior.}\label{table:n150_select} 
\end{table}
\end{center}

However, as is often the case with Bayesian methods, the influence of the prior is reduced as the sample size increases. In this case, the cost penalization from the FND prior diminishes as the sample size increases.  As a result, for large sample sizes, the cost-penalized selection using the FND prior more closely resembles the benefit-only model selection.  To view the changing impact of the cost penalization on selection as sample size increases, consider a similar data set of size $n=600$.  To mimic collection of additional data, we added $450$ new data points simulated from the settings in \eqref{eq:beta_vec} and \eqref{eq:cost_vec} to the set of $n=150$ studied in Table \ref{table:n150_select}, resulting in a data set of size $n=600$.  For this larger data set we again performed model selection first using a benefit-only analysis with uniform priors \eqref{eq:unif_priors} and then using cost-penalized Bayesian model selection with the FND prior on the model space.  Both selection approaches choose the same model with all $6$ predictors with non-null effect sizes, with a total cost of 26 minutes per observation.  Table \ref{table:n600_select} highlights the identical selection results from these two analyses for the size $n=600$ data set.  The C-statistic for the model containing every non-null predictor is 0.865, with posterior model probability equal to 0.594 from the benefit-only analysis (model prior 0.002) and cost-adjusted posterior model probability equal to 0.927 using the FND prior for selection (model prior 2e-29).

With the two selection results being identical, we can see that the cost penalization provided by the FND prior was ineffective after the addition of only $450$ more observations.  Thus, the reward for incurring the expense of additional data collection is to recommend a more expensive (or inefficient) model, the same as when no cost penalty is used.  If looking to decide which predictors to collect, for example, in a future study, a practitioner may use the FND prior and come to very different decisions based on the size of their existing data, by only a few hundred observations.  This points to a need to be able to adjust the cost penalization to provide cost-effective model selection for each medical application at hand.  Section \ref{subsec:sim_study_inclusion_path} demonstrates how the ECP can be used with our inclusion path approach to adjust the cost penalization so that it can be maintained at different sample sizes.

\begin{center}
\begin{table}[h!]
\centering
\begin{tabular}{ cc }   
Benefit-only selection & FND selection\\  
$n=600$  & $n=600$ \\  
\begin{tabular}{ |c|c|c|c| } 
 \hline
 \textbf{Predictor} & \textbf{Effect} & \textbf{Cost}& \textbf{Cost} \\ 
 & \textbf{size} & \textbf{level} &  \textbf{(mins.)}\\
 \hline
 $X_1$ & null & baseline & 1 \\ 
 \hline
 $X_2$ & null & cheap & 3 \\
 \hline
 $X_3$ & null & expensive & 9 \\
 \hline
 \rowcolor{Dandelion}
 $X_4$ & smaller & baseline & 1 \\ 
 \hline
 \rowcolor{Dandelion}
 $X_5$ & smaller & cheap & 3 \\   
 \hline
 \rowcolor{Dandelion}
 $X_6$ & smaller & expensive & 9 \\ 
 \hline
 \rowcolor{Dandelion}
 $X_7$ & larger & baseline & 1 \\ 
 \hline
 \rowcolor{Dandelion}
 $X_8$ & larger & cheap & 3 \\   
 \hline
 \rowcolor{Dandelion}
 $X_9$ & larger & expensive & 9 \\ 
 \hline
 \rowcolor{white}
 \multicolumn{3}{c}{} & \multicolumn{1}{c}{\underline{\textbf{26/obs}}} \\
\end{tabular} &  
\begin{tabular}{ |c|c|c|c| } 
 \hline
\textbf{predictor} & \textbf{Effect} & \textbf{Cost}& \textbf{Cost} \\ 
 & \textbf{size} & \textbf{level} & \textbf{(mins.)}\\ \hline
 $X_1$ & null & baseline & 1 \\ 
 \hline
 $X_2$ & null & cheap & 3 \\
 \hline
 $X_3$ & null & expensive & 9 \\
 \hline
 \rowcolor{Dandelion}
 $X_4$ & smaller & baseline & 1 \\ 
 \hline
 $X_5$ & smaller & cheap & 3 \\   
 \hline
 \rowcolor{Dandelion}
 $X_6$ & smaller & expensive & 9 \\ 
 \hline
 \rowcolor{Dandelion}
 $X_7$ & larger & baseline & 1 \\ 
 \hline
 \rowcolor{Dandelion}
 $X_8$ & larger & cheap & 3 \\   
 \hline
 \rowcolor{Dandelion}
 $X_9$ & larger & expensive & 9 \\ 
 \hline
 \rowcolor{white}
 \multicolumn{3}{c}{} & \multicolumn{1}{c}{\underline{\textbf{26/obs}}}\\
\end{tabular} \\
\end{tabular}
\caption{Candidate predictors and their corresponding effect sizes and costs. Rows colored in orange  \fcolorbox{black}{Dandelion}{\rule{0pt}{5pt}\rule{5pt}{0pt}}\quad are the predictors selected by (left) the benefit-only analysis and (right) cost-penalized selection using the FND prior for a data set of size $n=600$.  After the addition of only $450$ more observations to the data set studied in Table \ref{table:n150_select}, the benefit-only selection and Bayesian model selection using the FND prior choose the model with the same predictors and cost per observation, thus nullifying the default cost penalization from the FND prior.}\label{table:n600_select} 
\end{table}
\end{center}

\subsection{Inclusion paths using adjusted cost penalization} \label{subsec:sim_study_inclusion_path}

\begin{figure}[]
\centering
\begin{subfigure}[b]{0.5\textwidth}
  \centering
  \includegraphics[width=1\linewidth]{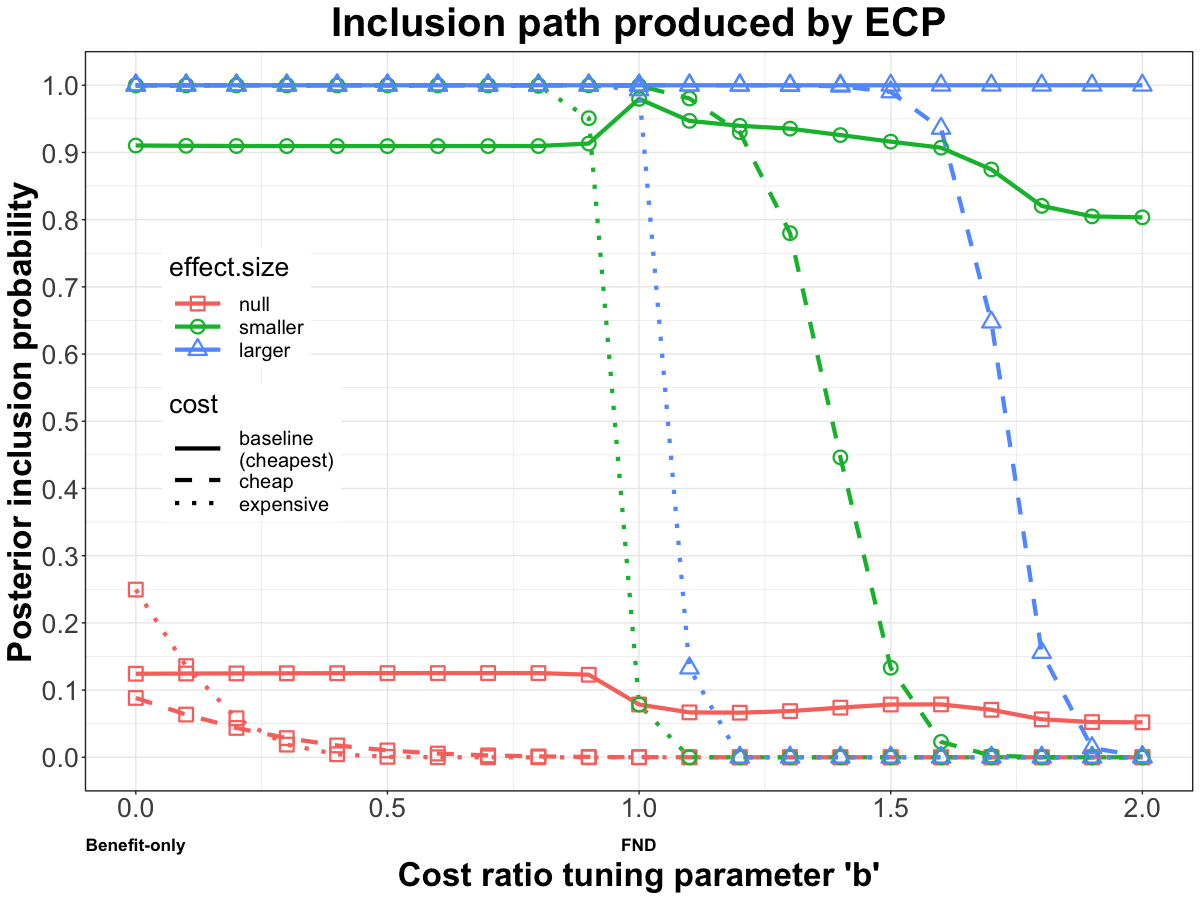}
  \caption{}
  \label{fig:inclusion}
\end{subfigure}
\begin{subfigure}[b]{0.5\textwidth}
  \centering
  \includegraphics[width=1\linewidth]{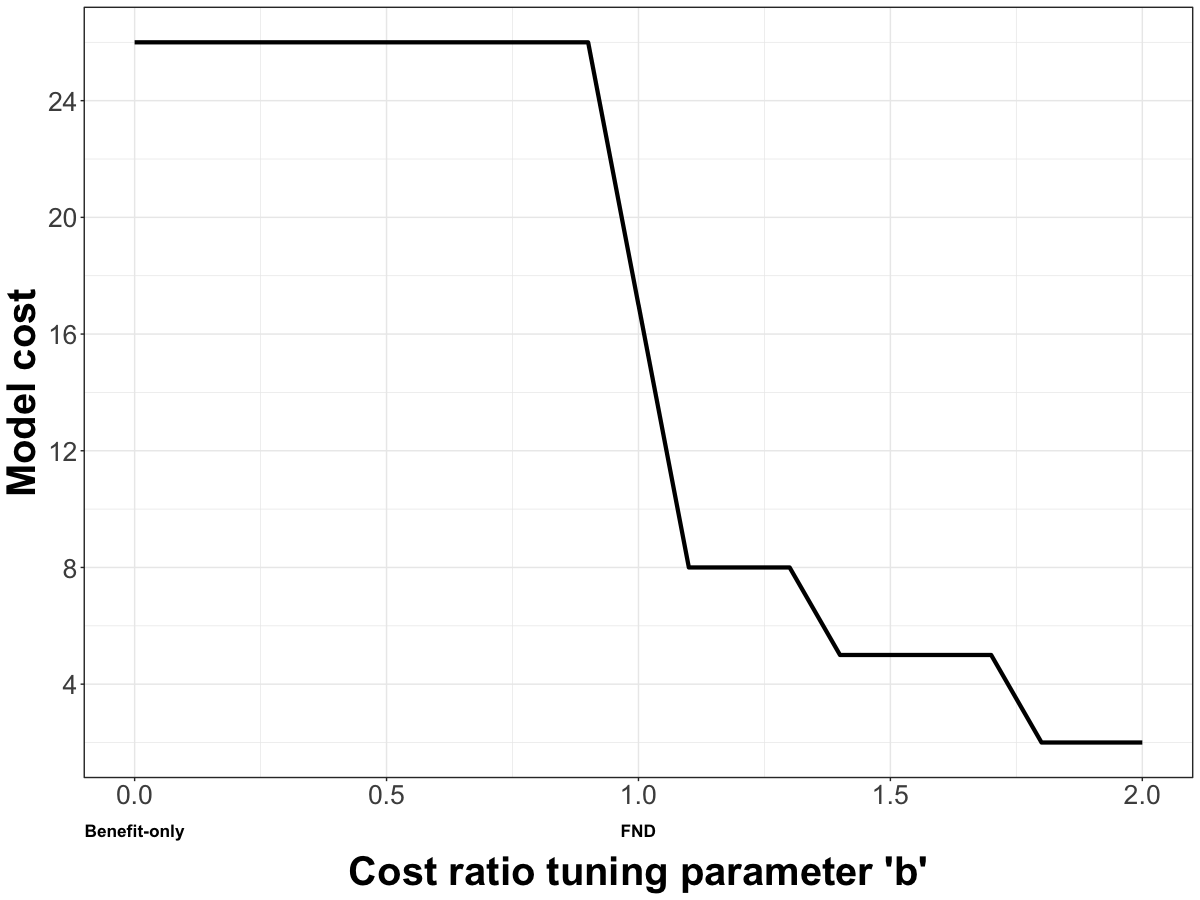}
  \caption{}
  \label{fig:cost}
\end{subfigure}
\begin{subfigure}[b]{0.5\textwidth}
  \centering
  \includegraphics[width=1\linewidth]{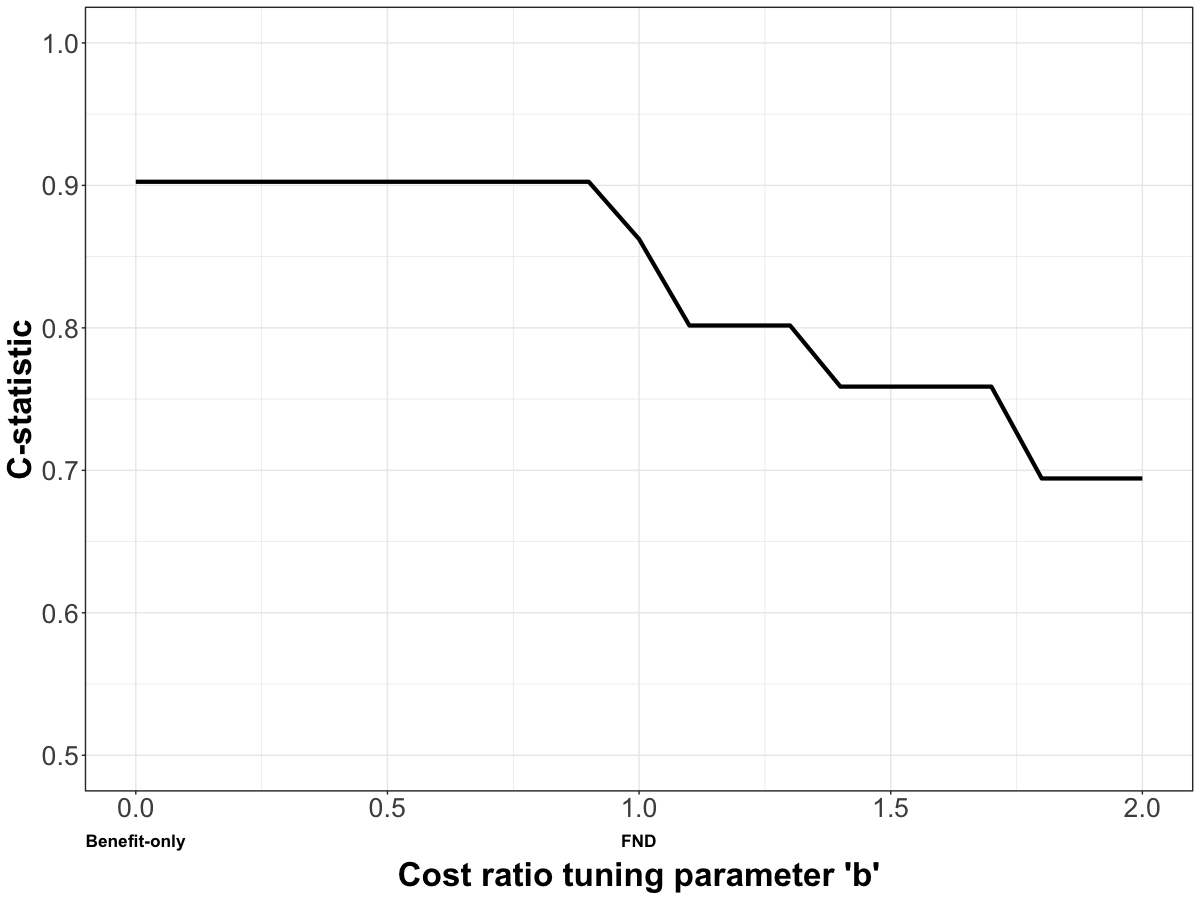}
  \caption{}
  \label{fig:cstat}
\end{subfigure}
\caption{(a) Inclusion paths for each of the $9$ predictors with baseline, cheap, and expensive costs and null, smaller, and larger effect sizes for the $n=450$ data set, (b) the cost of the selected model (MAP and median probability models are the same for this data set), and (c) the C-statistic of the selected model.  Model selection was performed using the ECP and tuning parameter values $b$ from 0 and 2 in increments of 0.1.  Values of $b$ above $2$ are not shown here because there is no further change in the posterior inclusion probabilities or selected models.}
\label{fig:3plots}
\end{figure}

Visualizing the effect of the tuning parameter $b$ enables the practitioner to see the change in all the predictors' importance in the posterior for a range of cost penalizations.  This forms the basis for our proposed inclusion path.  Here, we demonstrate the proposed inclusion path approach with the ECP assigned to the model space.  Consider a data set of size $n=450$, where the impact of the FND prior begins to diminish, simulated as described in Section \ref{subsec:sim_study}.  A similar example with correlated predictors appears in the Supplementary Material.  We apply the ECP with cost ratio function \eqref{eq:exp_adjust_formula} and a range of values for the tuning parameter $b$.  We apply each of the resulting ECP model priors to the model space and obtain posterior inclusion probabilities (benefit-only or cost-adjusted) for each of the $9$ candidate predictors.  Then we create an inclusion path by plotting the posterior inclusion probabilities (benefit-only or cost-adjusted) as a function of tuning parameter $b$.  Figure \ref{fig:inclusion} plots the inclusion path for each combination of candidate predictor cost and effect size when the ECP is used to adjust the amount of cost penalization.


From Figure \ref{fig:inclusion}, the posterior inclusion probabilities for $8$ of the $9$ predictors decrease as the cost penalization is increased via larger values of $b$. The predictor with the larger effect size and baseline cost maintains a posterior inclusion probability (benefit-only or cost-adjusted) at $1$ across all values of $b$, as it is not penalized \textit{a priori} by the FND or ECP priors, according to property (d) from Section \ref{subsec:inclusion_path_setup}.  The posterior inclusion probabilities for the cheap and expensive null predictors are 
$0.09$ and $0.25$, respectively, in the benefit-only analysis corresponding to $b=0$. The values of $0<b<1$ introduce cost penalization that is less severe than that from the FND prior but higher than the (nonexistent) cost penalization from the benefit-only analysis.  For example, the cost-adjusted posterior inclusion probabilities for the cheap and expensive null predictors drop to $0.04$ and $0.06$ when $b=0.2$ is used with the ECP. Then the null predictor with expensive cost has $0$ cost-adjusted posterior inclusion probability for $b \geq 0.7$.  Thus, even slightly penalizing these null predictors for having costs above the baseline helps to move their inclusion probabilities to $0$.  For $b=1$ (the FND approach), the null predictor with cheap cost has cost-adjusted posterior inclusion probability $0.0002$, which decreases to $0$ for $b \geq 1.1$.  The posterior inclusion probability of the null predictor with baseline cost is $0.12$ in the benefit-only analysis and the cost-adjusted posterior inclusion probability equals $0.05$ for $b \geq 3.5$, as this predictor has a null effect but is not penalized by any variation of the ECP.
When determining which predictors are selected with each variation of the ECP, we can consider comparing the cost-adjusted posterior inclusion probabilities for the remaining non-null predictors to a threshold, e.g.\@ $0.5$ as in the median probability model \citep{Barbieri2004}.  The inclusion path allows us to see which predictors' cost-adjusted inclusion probabilities meet the desired threshold for the different versions of cost-penalized selection at one time. For example, in the benefit-only analysis, which corresponds to the ECP with $b=0$, every one of the $6$ non-null predictors have posterior inclusion probabilities beyond the $0.5$ threshold under consideration.  Using the FND prior, which corresponds to the ECP with $b=1$, the cost-adjusted posterior inclusion probabilities for all non-null predictors except for the expensive one with smaller effect size exceed the $0.5$ threshold.  The FND approach is model selection consistent.  Thus, for our simulated examples where $p=9$ and the full model space can be enumerated, the FND approach selects the true generating model with non-null predictors as the sample size increases.  Figure \ref{fig:3plots} indicates that the cost-adjusted posterior inclusion probabilities fall below 0.5 at b = 1, 1.1, 1.5, and 1.8 for these predictors in the following order: expensive with smaller effect size, expensive with larger effect size, cheap with smaller effect size, and cheap with larger effect size.

The practitioner may choose a different threshold for the posterior inclusion probabilities (benefit-only or cost-adjusted) to suit their application.  The goal of the inclusion path is to provide a way to select the best model that conforms to the practitioner's budget by deciding which predictors to invest in.  We use $0.5$ for illustration here.

The cost-adjusted posterior inclusion probabilities for this data set decrease with $b$ first according to cost and then by effect size.  We found this trend to be true for this particular data set when a linear function of the cost ratio according to $b$ is used as well; see the Supplementary Material.  Now imagine that the practitioner is planning a study and wishes to refrain from collecting some of the predictors, either to lower the overall cost or to be able to record more total observations.  If the cost per observation for the benefit-only model here is too high, the practitioner might consider dropping/excluding predictors in an order first according to cost and then according to effect size, as indicated by the paths in Figure \ref{fig:inclusion}.  A visual such as that in Figures \ref{fig:inclusion} is important because it enables the practitioner to see how the different levels of cost penalization affect the posterior inclusion probabilities of individual predictors while also accounting for their effect sizes.

Figure \ref{fig:cost} plots the cost of the selected model for each value of $b$ used to create the inclusion path in Figure \ref{fig:inclusion}.  For this data set, the MAP model and median probability model are the same for all values of $b$.  We can see that the cost of the selected model drops each time a predictor is no longer in the model, e.g.\@ inclusion falls below 0.5.  Figure \ref{fig:cstat} plots the accuracy in the form of the C-statistic of the selected model for each value of $b$.  Similar to the cost in Figure \ref{fig:cost}, the C-statistic drops each time a predictor leaves the selected model as a result of a higher cost penalty.  A practitioner, patient, or other stakeholder may use Figures \ref{fig:cost} and \ref{fig:cstat} in conjunction with the inclusion path so they may keep any necessary budget, or desired efficiency or accuracy in mind when selecting cost-effective predictors.

\subsection{Case study: selecting cost-effective predictors to model diagnosis of heart disease} \label{subsec:heart_disease}
We apply our adjusted cost-penalizing model selection to a data set of clinical test results originally from \cite{clevelanddata}.  The data consists of the medical records of $n=297$ patients collected at the Cleveland Clinic Foundation in 1988. We use these data for illustration because diagnosis of medical conditions such as heart disease is an important classification problem and the data are available along with the financial cost of each candidate predictor.  We obtained these data from the UCI Machine Learning Repository.  The response is a binary variable that indicates the presence of heart disease in each patient; $Y_i=1$ if the patient has heart disease and $Y_i=0$ if the patient does not have heart disease, where $i=1, \ldots, n$.  There are $13$ candidate predictors.  A summary of numeric candidate predictors appears in Table \ref{tab:numeric_predictor} and categorical candidate predictors are summarized in Table \ref{tab:cat_predictor}.  Thus, there is a total of 8,192 models in the model space $\mathcal{M}$.  Each predictor has an associated cost per patient, listed in the third column of Tables \ref{tab:numeric_predictor} and \ref{tab:cat_predictor}.  The costs of the individual predictors range from $\$1.00$ to $\$102.90$, so the cost per observation ranges from $\$0.00$ (intercept-only) to $\$600.57$ for all 13 predictors per patient.  Costs listed in Tables \ref{tab:numeric_predictor} and \ref{tab:cat_predictor} are per patient and are specified in Canadian dollars, based on information from the Ontario Health Insurance Program.

\begin{center}
\begin{table}[h!]
\centering
\begin{tabular}{ |c|c|c|c|c|c| } 
 \hline
 \textbf{Predictor} & \textbf{Description} & \textbf{Cost} & \textbf{Mean} & \textbf{St.} & \textbf{Odds} \\ 
  \textbf{name} & & \textbf{(in \$)}& & \textbf{dev.} & \textbf{ratio*} \\
 \hline
 age & Patient age (years) & 1 & 54.54 & 9.05 & 1.61 \\ 
  \hline
 resting BP & Resting blood pressure (mm Hg) & 1 & 131.69 & 17.76 &  1.37 \\
  \hline
 cholesterol & Serum cholestoral (mg/dl) & 7.27 & 247.35 & 51.99 & 1.18 \\
  \hline
 heart rate & Maximum heart rate achieved & 102.90 & 149.60 & 22.94 & 0.36 \\
  \hline
 ST depression & ST depression induced by exercise & 87.30 & 1.06 & 1.17 & 2.91 \\
 & relative to rest & & & & \\
  \hline
\end{tabular}
\caption{Numeric candidate predictors for the Cleveland heart disease data. *The odds ratios are expressed in terms of an increase of one standard deviation in the predictor, and reflect the odds of the patient having a positive heart disease diagnosis.} \label{tab:numeric_predictor}
\end{table}
\end{center}

\begin{center}
\begin{table}[h!]
\centering
\begin{tabular}{ |c|c|c|c|c|c| } 
 \hline
 \textbf{Predictor} & \textbf{Description} & \textbf{Cost} & \textbf{Values} & \textbf{Percent} & \textbf{Odds} \\ 
 \textbf{name} & & \textbf{(in \$)} & & \textbf{observed} & \textbf{ratio**} \\
 \hline
 sex & Patient sex & 1 & 0-Female & 32.3\% & - \\ 
 & & & 1-Male & 67.7\% & 3.57 \\
 \hline
 blood & Fasting blood  & 5.20 & 0-False & 85.5\% & - \\
 sugar & sugar & & 1-True & 14.5\% & 1.02 \\
 \hline
 exercise & Exercise-induced & 87.30 & 0-No & 67.3\% & - \\
 angina & angina & & 1-Yes & 32.7\% & 7.00 \\
\hline
 chest & Chest pain & 1 &1- Typical angina & 7.7\% & - \\
 pain & type & & 2-Atypical angina & 16.5\% & 0.51 \\
 & & & 3-Non-anginal pain & 27.9\% & 0.63 \\
 & & & 4-Asymptomatic & 47.8\% & 6.04 \\
 \hline
 EKG & Resting & 15.50 &0-Normal & 49.5\% & - \\
 & electrocardiogram  & & 1-ST-T wave abnormality & 1.3\% & 5.02 \\
 & results & & 2-Probable/definite left & 49.2\% & 1.97 \\
 \hline
 peak ST & Slope of the & 87.30 & 1-Upsloping & 46.8\% & - \\
 segment & peak exercise & & 2-Flat & 46.1\% & 5.31 \\
 & ST segment & & 3-Downsloping & 7.1\% & 3.81 \\
 \hline
 major & Major vessels & 100.90 & 0 & 58.6\% & - \\
 vessels & colored by & & 1 & 21.9\% & 6.01 \\
 & flourosopy & & 2 & 12.8\% & 12.70 \\
 & & & 3 & 6.7\% & 16.24 \\
 \hline
 defect & Type of heart  & 102.90 & 3-Normal defect & 55.2\% & - \\
 type & defect & & 6-Fixed defect & 6.1\% & 6.87 \\
 & & & 7-Reversible defect & 38.7\% & 11.19 \\
 \hline
\end{tabular}
\caption{Categorical candidate predictors for the Cleveland heart disease data.  **Odds ratios measure the shift in multiplicative odds from the reference category (denoted by -) of having a positive heart disease diagnosis.} \label{tab:cat_predictor}
\end{table}
\end{center}

\begin{figure}[h!]
\centering
  \includegraphics[width=0.85\linewidth]{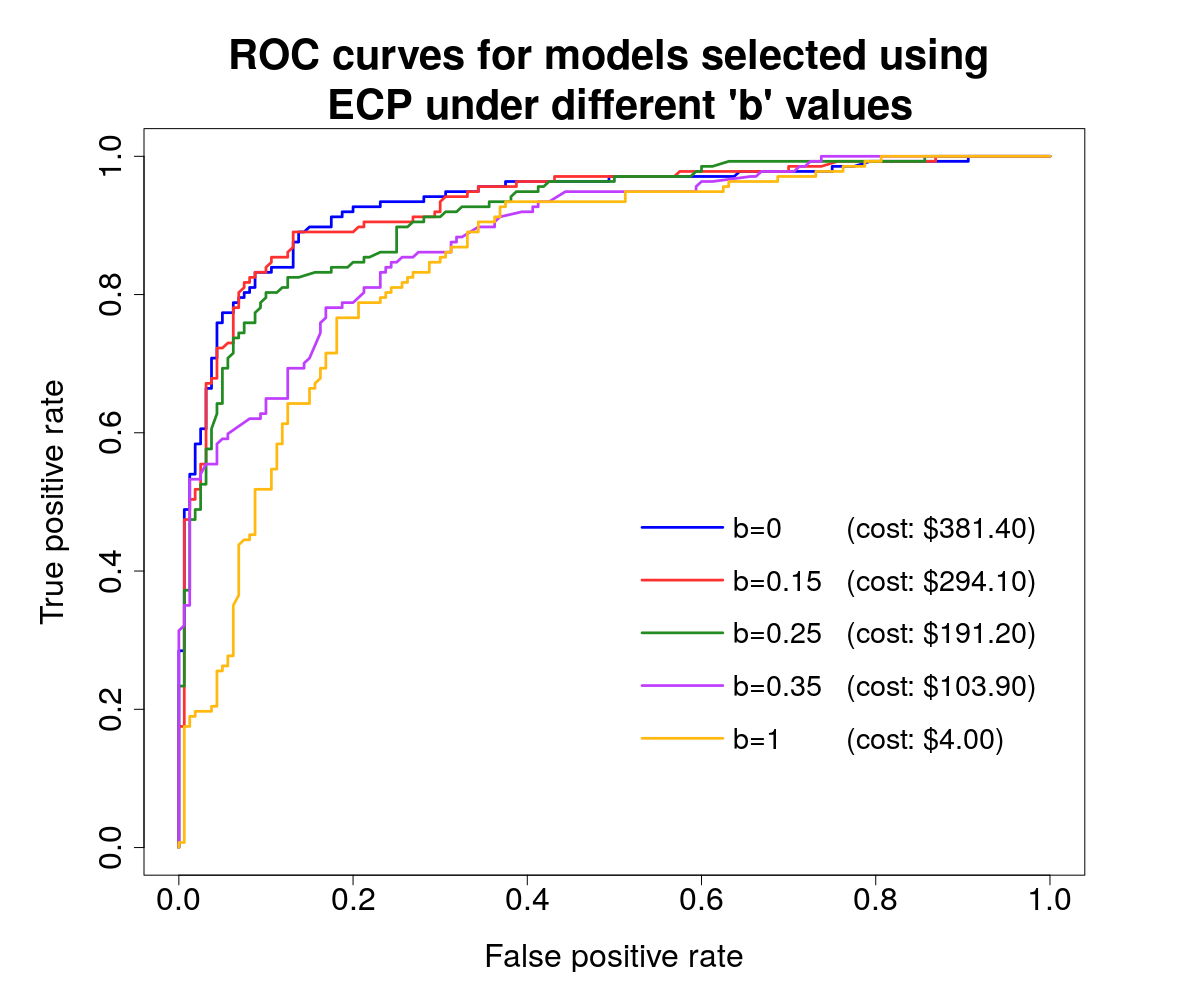}
\caption{ROC curves for median probability models for the Cleveland heart disease data using Bayesian model selection with the ECP model prior and different values of tuning parameter $b$.  The median probability and the MAP model for $b \geq 0.45$ is always the cost \$4.00 model containing only the four baseline predictors.  $b=1$ corresponds to the FND model prior.}
\label{fig:ROC_curves}
\end{figure}

We performed Bayesian model selection using the heart disease data, applying a benefit-only approach and cost-penalized model selection as described in Section \ref{sec:data_methods} with the FND and ECP priors assigned to the model space.  Benefit-only model selection with uniform priors \eqref{eq:unif_priors} on the model space selects the median probability model with the following predictors: sex, chest pain, resting BP, ST depression, peak ST segment, major vessels, and defect type with corresponding posterior model probability $0.072$, up from a prior model probability equal to 0.000122 and with a total cost of $\$381.40$ per patient.  Model selection using the FND prior on the model space selects the median probability model with the four baseline predictors (sex, age, chest pain, and resting BP), with corresponding cost-adjusted posterior model probability equal to 0.5.  This model that contains only baseline predictors has prior model probability equal to $0.063$ using the FND prior, and a cost equal to $\$4.00$ per patient.  While cost-penalized selection with the FND prior leads to a less costly model that is only $2.3\%$ of the cost of the benefit-only model, this model performs worse than more costly models in terms of classification.  Figure \ref{fig:ROC_curves} displays ROC curves and corresponding costs for models selected using different values of $b$ with ECP; the values of $b$ listed correspond to the median probability model.  Detection of a health condition such as heart disease is critical for doctors to be able to provide effective treatment and advice to affected patients. The cost $\$4.00$ model might not have acceptable false positive rates for the hospital to trust the diagnosis, and thus it is desirable to be able to adjust the cost penalization, particularly between $0<b<1$, for these data.  Our method is key for striking a balance between the benefit-only approach and the FND approach, by furnishing a spectrum of cost penalization choices.  The C-statistics for the models considered in Figure \ref{fig:ROC_curves} are 0.934, 0.931, 0.917, 0.887, and 0.849 for the median probability models from the ECP with $b=$ 0, 0.15, 0.25, 0.35, and 1, respectively.  To assess out-of-sample predictive accuracy, we also calculated an average C-statistic based on 10-fold cross-validation for each model by obtaining predicted probabilities for each observation based on regression coefficient estimates obtained without each of the 10 folds; these values were 0.914, 0.913, 0.902, 0.86, and 0.833, respectively.  See the Supplementary Material for more details.  We can see that penalizing based on predictor costs (e.g.\@ with $b=0.15$) can lead to great cost reduction with very little loss in performance compared to the benefit-only model.  Perhaps surprisingly, the model with cost equal to only $\$4.00$ per patient has a C-statistic of 0.849, which indicates that it might have some utility in a triage scenario.  However, our method provides the ability for practitioners to see which predictors can supplement the model to improve diagnosis beyond easily-obtained predictors like age and sex and a subjective pain rating from the patient.  

\begin{figure}[]
\centering
\begin{subfigure}[b]{0.5\textwidth}
  \centering
  \includegraphics[width=1\linewidth]{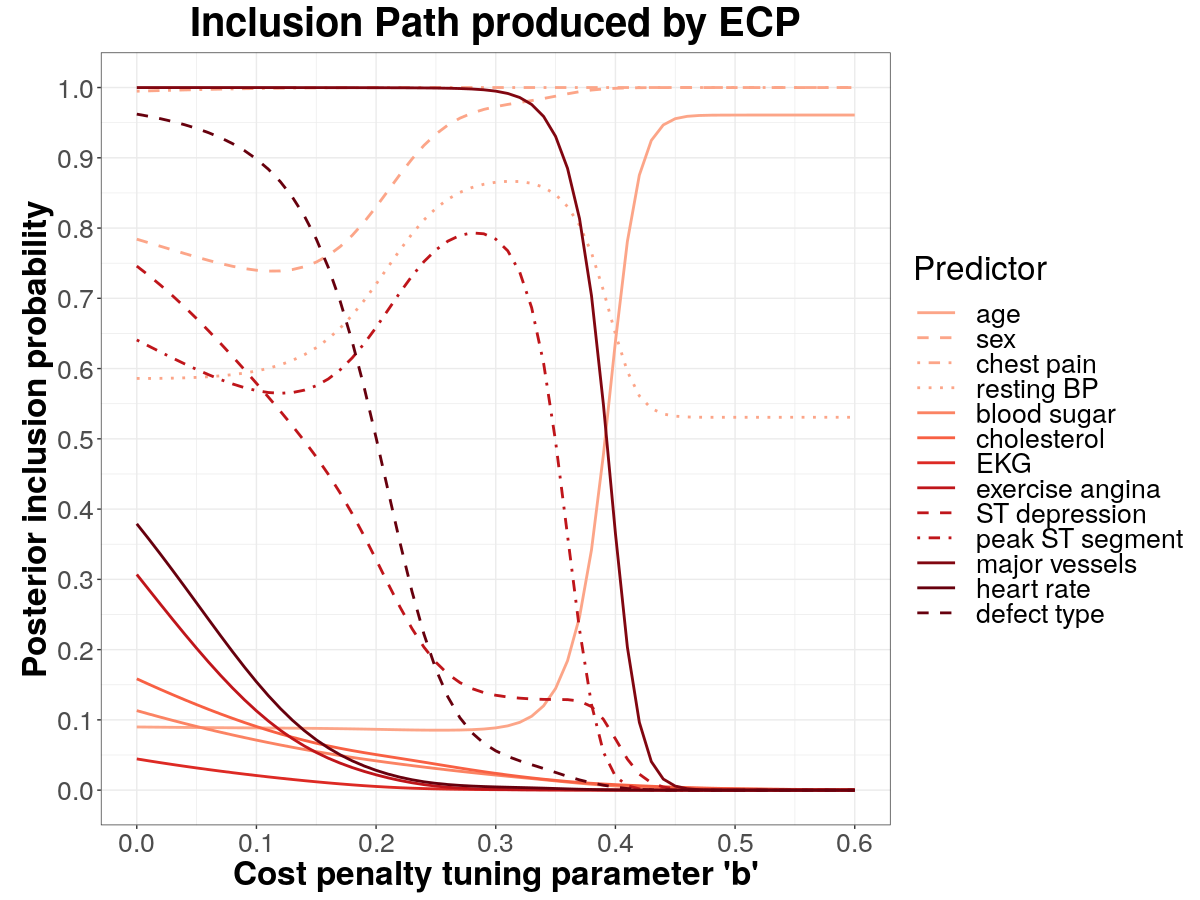}
  \caption{}
  \label{fig:heart_inclusion}
\end{subfigure}
\begin{subfigure}[b]{0.5\textwidth}
  \centering
  \includegraphics[width=1\linewidth]{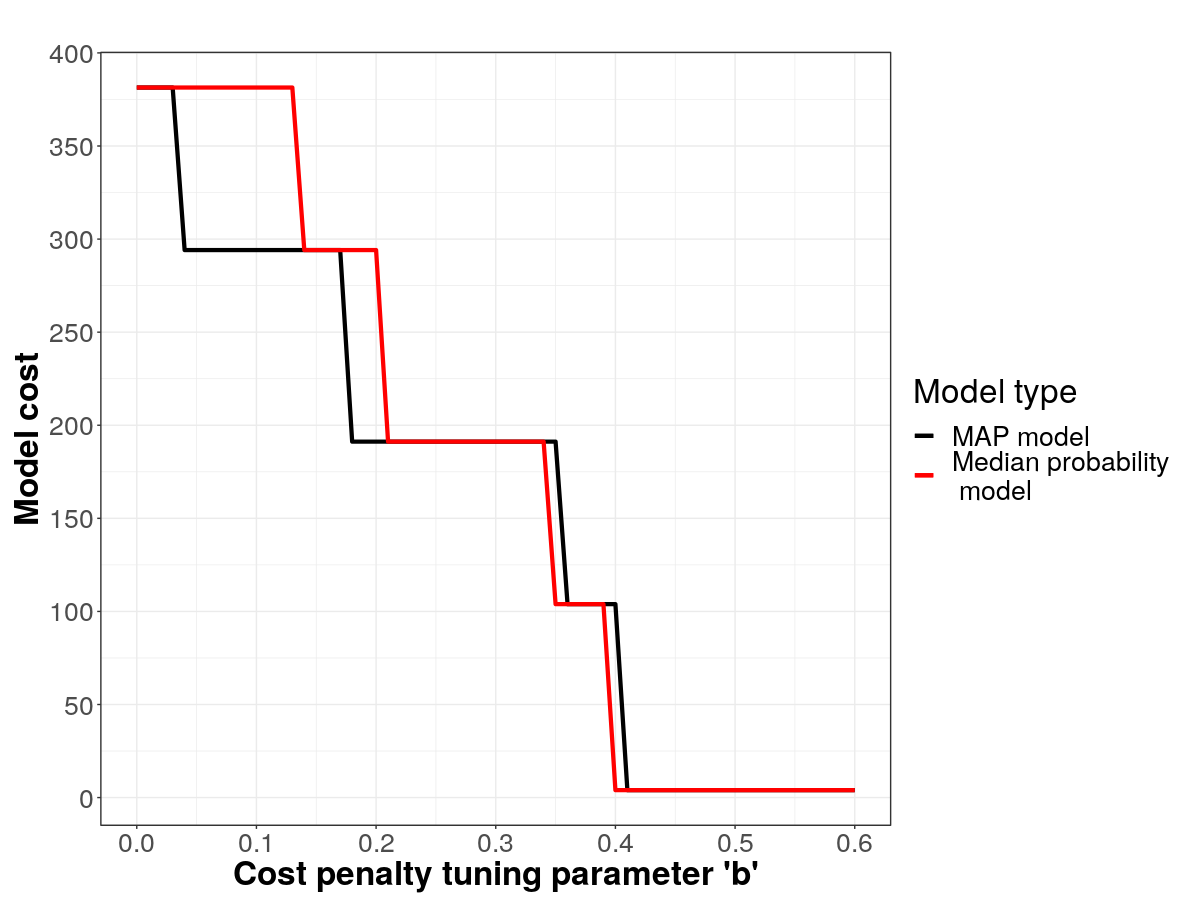}
  \caption{}
  \label{fig:heart_cost}
\end{subfigure}
\begin{subfigure}[b]{0.5\textwidth}
  \centering
  \includegraphics[width=1\linewidth]{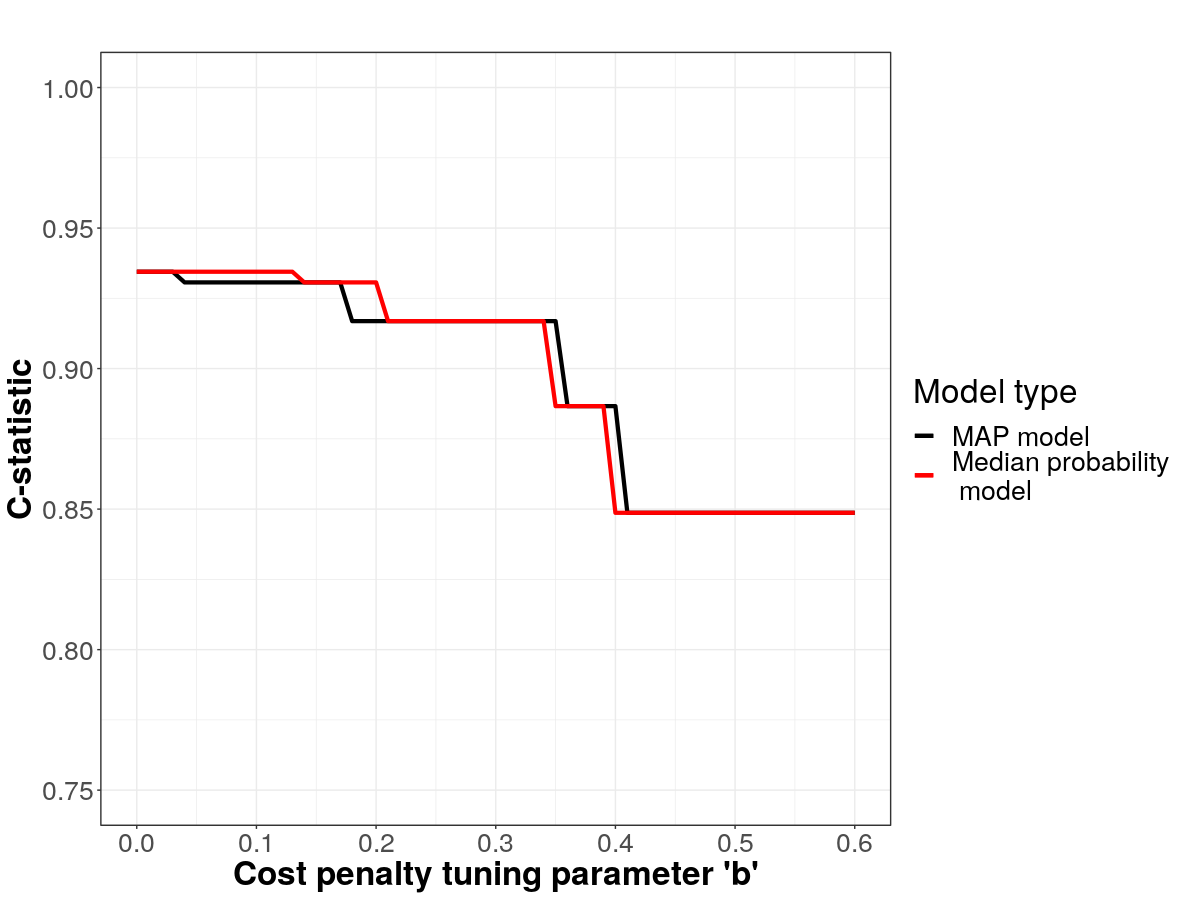}
  \caption{}
  \label{fig:heart_cstat}
\end{subfigure}
\caption{(a) Inclusion paths for each of the $13$ predictors from the heart disease data, (b) the cost of the MAP and median probability models, and (c) the C-statistic of the MAP and median probability models.  Model selection was performed using the ECP and tuning parameter values $b$ between 0 and 0.6.  Values of $b$ above $0.6$ are not shown here because there is no further change in the posterior inclusion probabilities or selected models.}
\label{fig:heart_3plots}
\end{figure}

To study the individual candidate predictors for diagnosing heart disease, we applied our inclusion path approach with adjusted cost penalization to the heart disease data from \cite{clevelanddata}.  Figure \ref{fig:heart_inclusion} contains the inclusion path using the ECP with tuning parameter values of $b$ between 0 and 0.6.  The lines corresponding to each predictor's inclusion path are color-coded such that lines corresponding to predictors with the same cost are the same color; darker shades of red correspond to more costly predictors, and then line types differ between distinct predictors that have the same cost.  Values of $b$ above $0.6$ are not shown here, as the cost-adjusted posterior inclusion probabilities do not change for $b > 0.6$ for these data, and the cost $\$4.00$ model containing only the baseline predictors is selected with highest posterior model probability for all $b > 0.5$.  

From Figure \ref{fig:heart_inclusion} we can see that the posterior inclusion probabilities for the baseline predictors sex and chest pain are high for the benefit-only analysis as well as all cost-penalized analyses, with the cost-adjusted posterior inclusion probabilities being equal to $1$ for all values of $b \geq 0.4$.  The posterior inclusion probability for baseline predictor age is only $0.09$ in the benefit-only analysis; then the cost-adjusted posterior inclusion probability increases to $0.64$ at $b=0.4$ and reaches and remains at $0.96$ for $b \geq 0.5$.  The final baseline predictor, resting BP, has posterior model probability equal to $0.59$ in the benefit-only analysis.  Its cost-adjusted posterior inclusion probability increases with $b$ until $b=0.3$, then decreasing to $0.53$ for $b \geq 0.1$, always remaining above a $0.5$ probability threshold.  Predictors heart rate, exercise angina, cholesterol, EKG, and blood sugar have posterior inclusion probabilities below $0.5$ in the benefit-only analysis; cost-adjusted posterior inclusion probabilities for these five predictors all move towards $0$ as the cost penalty increases, indicating that these predictors would not be chosen based on a benefit-only or a cost-penalized analysis unless a lower threshold is used.  The posterior inclusion probabilities for ST depression and defect type are equal to $0.75$ and $0.96$, respectively, in the benefit-only analysis.  The cost-adjusted posterior inclusion probabilities for both ST depression and defect type continue to decrease towards 0 as $b$ increases.  Finally, the peak ST segment predictor has posterior inclusion probability 0.64 for $b=0$; its cost-adjusted posterior inclusion probability fluctuates until $b=0.3$ and then decreases towards 0.


Since we saw from Figure \ref{fig:ROC_curves} that the cursory, baseline predictors included at $b=1$ may not be ideal by themselves, we might think of sliding $b$ below $1$ until reaching a set of predictors that provide a cost-penalized model whose performance is adequate and whose overall cost satisfies a hospital or patient budget.  Based on a $0.5$ probability threshold, a hospital might consider, in addition to the four baseline predictors that are easy to obtain, also including predictors such as major vessels, peak ST segment, defect type, and ST depression.  Suppose, for example, that a particular hospital has a maximum budget of $\$200.00$ per patient for the purpose of diagnosing heart disease.  The hospital could recommend its providers record the four baseline predictors for each patient, as well as the major vessels and peak ST segment predictors, for a total cost of $\$192.20$ per patient.  The inclusion paths make it clear which predictors would be included/excluded from selection at each level of cost penalization, and can be used to visualize inclusion for several predictors in any medical setting with quantifiable costs over a wide range of cost penalizations.

It is important to note that costs are often subjective, and who controls or specifies the $c_j$'s for the model prior is of great importance.  In Section \ref{subsec:heart_disease}, we have used the monetary costs that were provided for this particular case study data.  While the costs that are known for this case study pertain to the monetary cost to hospital and insurance stakeholders, in other applications a patient could also settle on their own set of costs based on comfort, risk, convenience, etc. and apply our proposed method.  Perhaps patient advocates or medical practitioners could subjectively elicit values for the $c_j$’s in a useful way.  A primary challenge to using our method would be to have patients or other stakeholders put forth their values for the individual $c_j$’s to capture/quantify the relative cost of the candidate predictors to that user.

\subsection{Computation}
All computations for our proposed method were performed in R.  We use a Laplace approximation to approximate each of the integrated likelihoods. Specifically, for each candidate model, we use \texttt{optim()} in R to obtain the posterior mode of the regression coefficients and the Hessian matrix evaluated at the posterior mode by maximizing over a function containing the joint distribution of the data and regression coefficients $\boldsymbol{\beta}_{\boldsymbol{\gamma}}$.  Given responses $\textbf{Y}$, a matrix $X$ containing all candidate covariates, a vector of marginal costs for each candidate covariate, and a specific value for $b$, we do the following: (1) compute the set of integrated likelihoods for every candidate model using Laplace approximation, (2) form Bayes factors as in Equation \eqref{eq:standard_bf} for each candidate model with respect to a baseline model, (3) calculate the model prior $P(\boldsymbol{\gamma})$ for each candidate model as in Equation \eqref{eq:adjusted_prior}, and (4) calculate the set of $2^p$ posterior model probabilities (benefit-only or cost-adjusted) as in Equation \eqref{eq:post_prob}.

Computations for the case study in Section \ref{subsec:heart_disease} were completed using a $2 \times \text{E}5 - 2683\text{v}42.1$GHz (Broadwell) CPU supercomputer from Advanced Research Computing at Virginia Tech.  For a single value of $b$, it takes 16 minutes and 40 seconds to enumerate all models and calculate posterior model probabilities (benefit-only or cost-adjusted) for all $2^{13}$ candidate models.  The examples studied here contain relatively small numbers of candidate predictors, so the entire model space of size $2^p$ can be enumerated.  In the case of large $p$, our method could be adapted for use with a stochastic search algorithm to feasibly explore the model space, such as genetic algorithms \citep{wu2020hyper} or reversible jump \citep{Fouskakis2009, rjmcmc}.

\color{black}

\section{Discussion} \label{sec:discussion}

We have presented an approach to adjust cost-penalizing model priors for cost-adjusted Bayesian model selection.  Our approach extends the well-established FND prior \citep{Fouskakis2009} by giving the practitioner the ability to adjust the level of cost penalization on candidate predictors and visualize the resulting cost-adjusted posterior inclusion probabilities.  We proposed functions, according to a tuning parameter $b$, of the ratios of marginal predictor costs relative to a baseline cost.  The resulting ECP prior that we study provides adjusted levels of cost penalization controlled by the practitioner via the tuning parameter. The properties that our cost ratio functions adhere to ensure that our adjusted priors can easily reproduce a benefit-only analysis and allow for unit conversion without changing model selection results, making it useful for costs measured in a variety of ways.  We have shown, through simulation, that adjusting the cost penalty according to our proposed functions helps to maintain the cost penalization for larger sample sizes.  Our inclusion path approach, which plots the change in cost-adjusted posterior inclusion probabilities together across a range of adjusted cost penalties, provides a visual tool to learn the relative importance of predictors when accounting for their costs and to make decisions about individual predictors to meet an overall budget. Our method can be applied to any binary outcome (e.g. diagnosis) where medical practitioners need to make a decision or prediction based on predictors with quantifiable costs.  Our proposed method does not have all of the same theoretical properties derived in \cite{Fouskakis2009}; for example, our proposed model prior is not model selection consistent.  Rather, the purpose of our method is to enable the user to explore different cost penalty scenarios.

This work extends the utility of the FND prior by adjusting the penalization on costly predictors in model selection. For example, suppose that the model selected using the FND prior has a total predictor cost that exceeds the practitioner's or hospital's designated budget \citep{rjmcmc}.  Then our proposed inclusion path can easily be used to see which predictor(s) have probabilities that fall below the desired threshold as the practitioner slides $b$ towards higher values.  Similarly, if the practitioner seeks a higher-performing model that still penalizes candidate predictor costs to some extent, they can slide $b$ down, closer to $0$, to learn which additional predictor(s) will improve model fit without causing an undue increase in the cost per observation.  We applied our approach to a data set of $297$ heart disease patients and found that decreasing the cost penalization from the FND prior helps to identify predictors that can improve the diagnosis of heart disease while still appropriately penalizing the most costly predictors.  The ECP provides a useful option for adjusting the magnitude of cost penalization.  It is possible that, due to the presence of collinearity, large $p$, or other data characteristics, Bayesian model selection when the sample size is large may select different benefit-only and cost-penalized models with no adjustment to the cost penalty.  For example, \cite{Fouskakis2009} studied pneumonia-related deaths in $n=2,532$ patients with 83 candidate predictors, and cost-adjusted BIC selected a model which resulted in a 66.7\% reduction in cost compared to the benefit-only model.  Our method is intended to provide an extension for applications and data sets where the sample size impacts cost-penalized selection decisions, and to allow practitioners to consider how influential the prior should be with respect to their aim.

The matter of cost can implicitly raise ethical concerns, especially in the case of medical applications that directly impact the health and well-being of patients.  Medical practitioners would be well-positioned to apply our methods, as they have the knowledge and experience to understand how collecting different predictors may impact patients' health, comfort, and finances and can advise on setting the cost ratio tuning parameter with these considerations in mind.  We envision practitioners using a range of $g(\cdot)$ and $b$ values to study different cost penalizing settings and choose a model that best suits their budget and concerns.  We recommend that practitioners and researchers apply our cost-penalized selection methods in an ethical and transparent manner and acknowledge any implications of analyzing a particular type of cost.

  Avenues for future research include developing cost-penalized model selection for a sequential decision-making framework, for example, in medical diagnoses where practitioners may order additional tests depending on a patient’s initial results.  Investigation may also be done to find and recommend an upper bound for $b$ given a particular data set.  Our proposed method may also be adapted and applied more broadly to non-binary response data, for example to other generalized linear models.  Another possible extension could include adjusting the model space and/or prior to accommodate cost structures for grouped or discounted predictors.  Future development of an MCMC algorithm would allow for study of additional metrics of predictive ability based on the posterior distribution.  Another opportunity for exploration may include changes to the prior and selection for categorical predictors such as those in Table \ref{tab:cat_predictor}.  \cite{Gonzalo2022} find that selecting either all or none of the levels of a categorical predictor can have disadvantages such as understating the effect of a categorical predictor with many levels if only a few of those levels are relevant.  \cite{Gonzalo2022} propose to allow for selection of only active levels of a categorical predictor and propose a model prior which maintains $1/2$ prior inclusion for categorical variables regardless of the number of levels.  Further investigation is needed to understand how to apply this approach while incorporating a cost penalty on predictors through the model prior.
  
  Together, our tuning parameter-based functions of the cost ratios and our inclusion path proposed in this manuscript give the practitioner considerable flexibility to weigh each predictor's cost with its modeling ability, with probabilities providing a concrete measure of inclusion, especially relative to other predictors.

\section*{Acknowledgments}
We are thankful to Leidos for providing the funding for this work.  Computations for this manuscript have been performed on supercomputers of Advanced Research Computing at Virginia Tech.

\bibliography{ref}%




\end{document}


\begin{center}
{\Large \textbf{Supplementary Material for ``Flexible cost-penalized Bayesian model selection: developing inclusion paths with an application to diagnosis of heart disease"}}
\end{center}

\section{Useful definitions}
\textbf{KL divergence}\
    
    \noindent KL divergence is a divergence value that measures the difference between two probability distributions.  For two probability distributions $P$ and $Q$ with sample space $\mathcal{X}$, the KL divergence is calculated as:

\begin{equation}
        KL(P||Q) = \sum_{x \in \mathcal{X}} P(x) \log\Big(\frac{P(x)}{Q(x)}\Big). \tag{S.1}
    \end{equation}

Smaller values of $KL(P||Q)$ indicate that the two distributions are more similar to each other.  In Section 1 of the manuscript, we calculate the KL divergence between the posterior model probabilities produced by a benefit-only selection approach and a cost-penalized selection approach with the FND prior on the model space.  \\

\vspace{0.1cm}
\noindent \textbf{Receiver operating characteristic (ROC) curve}

\noindent The ROC curve is a plot of the performance of a binary classification model.  The ROC curve plots the false positive rate on the x-axis and the true positive rate on the y-axis at different decision thresholds.  \\

\vspace{0.1cm}
\noindent \textbf{Concordance (C) statistic}

\noindent The C-statistic is the area under the ROC curve for a binary classification model.  Higher values indicate better performance, and a C-statistic equal to 0.5 indicates that the binary classifier is no better at accurately predicting an outcome than random chance.

\vspace{0.1cm}
\noindent \textbf{10-fold cross-validation C-statistic}

\noindent In Section 3.4 of the manuscript, we assess the out-of-sample predictive accuracy of the different models selected for the heart disease data as $b$ is increased.  To assess this, we use 10-fold cross-validation to obtain a set of 10 predictive C-statistics for each of the 5 models selected for the heart disease data at different values of $b$.  Namely, for a given model $\boldsymbol{\gamma}$, we do the following:

    \begin{enumerate}
        \item  Randomly assign each of the $n=297$ observations to one of 10 folds, obtaining \{$\boldsymbol{Y}_F$, $\boldsymbol{X}_{\boldsymbol{\gamma},F}$\} for $F=1, \ldots, 10$.
        \item For a single fold, $F$, maximize the log of the joint distribution of the data and the prior on the regression coefficients using data from the remaining folds, i.e. $f(\boldsymbol{\beta_{\boldsymbol{\gamma}}}, \boldsymbol{Y}_{-F}|\boldsymbol{X}_{\boldsymbol{\gamma},-F})$ using \texttt{optim()} in R.
        \item Denote the regression coefficients obtained from step 2 as $\hat{\boldsymbol{\beta}}_{\boldsymbol{\gamma},-F}$.  Obtain the predicted logit values for the observations in fold $F$ as $\text{logit}(\hat{\boldsymbol{p}}) = \boldsymbol{X}_{\boldsymbol{\gamma},F} \hat{\boldsymbol{\beta}}_{\boldsymbol{\gamma},-F}$.  Obtain the predicted probabilities for the observations in fold $F$ as $\hat{\boldsymbol{p}} = \exp(\text{logit}(\hat{\boldsymbol{p}}))/(1+\exp(\text{logit}(\hat{\boldsymbol{p}})))$.  
        \item Obtain the predictive C-statistic for fold $F$ as the area under the ROC curve created from $\hat{\boldsymbol{p}}$ and $\boldsymbol{Y}_F$.
        \item Repeat steps 1-4, withholding each of the 10 folds once, to obtain 10 different predictive C-statistics for model $\boldsymbol{\gamma}$.
    \end{enumerate}

Figure \ref{fig:10fold_cstats} contains a plot of the resulting means and standard deviations for each model across the folds.  

\renewcommand{\thefigure}{S.\arabic{figure}}
\begin{figure}[h!]
  \centering
  \includegraphics[width=0.9\linewidth]{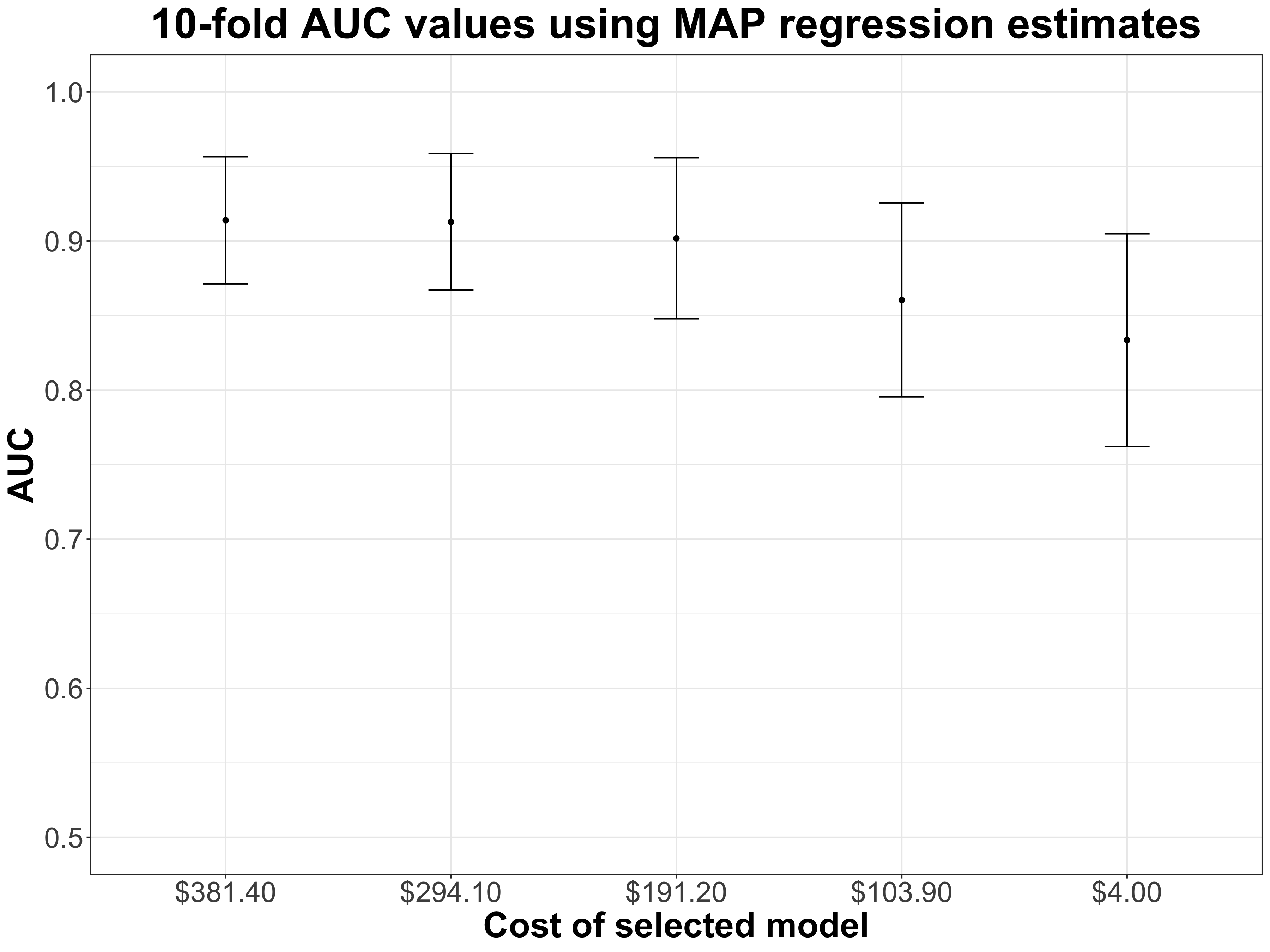}
  \caption{}
\label{fig:10fold_cstats}
\end{figure}

\color{black}
\section{Inclusion path algorithm}
Algorithm \ref{algo:inclusion_path} details the process for creating an inclusion path for data \textbf{Y} and $X_j, \ j=1,\ldots,p$ with the user's chosen function $g(\cdot)$ (e.g. exponential or linear), when $p$ is small enough to enumerate the entire model space.

\floatname{algorithm}{Algorithm}

\begin{algorithm}
\caption{Inclusion path procedure}\label{algo:inclusion_path}
\begin{algorithmic}[1]
\State Specify a grid of values for tuning parameter $b\geq 0$, say of length $B$.
\State for $i$ in \ 1:$B$
\begin{flushleft}
\{
\end{flushleft}
\State \hskip1.0em Compute $g(c_j/c_0,b) \ \forall j=1,\ldots,p$.
\State \hskip1.0em for $\ell \in \mathcal{M}$ 
\begin{flushleft}
\hskip2.5em \{
\end{flushleft}
\State \hskip3.5em Calculate $$P(M_\ell) = \exp\bigg\{-\frac{1}{2}\log n \sum_{j=1}^p \gamma_j \bigg(g\bigg(\frac{c_j}{c_0},b\bigg)-1\bigg)-\sum_{j=1}^p \log \big[n^{-\frac{1}{2}\big(g\big(\frac{c_j}{c_0},b\big)-1\big)}+1\big]\bigg\}$$
\State \hskip3.5em Approximate $p(\textbf{Y}|M_\ell)$ via Laplace approximation.
\State \hskip3.5em Calculate $$ P(M_{\ell}|\textbf{Y}) = \frac{p(\textbf{Y}|M_{\ell})P(M_{\ell})}{\sum_z{k=1}^K p(\textbf{Y}|M_k)P(M_k)}$$ 
\begin{flushleft}
\hskip2.5em \}
\end{flushleft}

    \State \hskip2.5em for $j$ in 1:$p$ (i.e., for each candidate predictor)
\begin{flushleft}
\hskip2.5em \{
\end{flushleft}
\State \hskip3.5em Calculate the posterior inclusion probability (benefit-only or cost-adjusted)
\begin{center}
    $P(\gamma_j=1|\textbf{Y}) = \sum_{\boldsymbol{\gamma}_k \in \mathcal{M}|\gamma_j=1} P(\boldsymbol{\gamma}_k|\textbf{Y}).$
\end{center}
\State \hskip3.5em Plot $P(\gamma_j=1|\textbf{Y}), \ j=1,\ldots,p$ at the current value of $b$ 
\begin{flushleft}
\hskip2.5em \}
\end{flushleft}

\begin{flushleft}
\}
\end{flushleft}
\end{algorithmic}
\end{algorithm}

\section{Additional simulation results}

\subsection{Linear Cost Prior (LCP)}
As noted in Section 2.3 of the manuscript, monotone functions of the cost ratio and tuning parameter $b$ are well-suited for the model prior in Equation (9).  

An alternative to the ECP is the linear cost prior (LCP). 
 For a given value of the tuning parameter $b$, the LCP takes the form of Equation (10), where $g(c_j/c_0,b)$ is a linear function of the cost ratio for predictor $X_j$:

\begin{equation} \label{eq:linear_adjust_formula}
 g\bigg(\frac{c_j}{c_0}\bigg) = \frac{(c_j - c_0) \cdot b + c_0}{c_0}. \tag{S.2}
\end{equation}

Note that \eqref{eq:linear_adjust_formula} satisfies properties (a)-(d) listed in Section 2.3 of the manuscript.  Figure S.1 contains plots for the inclusion path and cost and C-statistic of the selected models using the LCP on the same data set that was analyzed in Figure 3 of the manuscript using the ECP.  For this simulated data set, the order in which the cost-adjusted posterior inclusion probabilities for non-baseline predictors decrease towards 0 is the same as when the ECP was used.  Namely, as $b$ increases, the cost-adjusted posterior inclusion probabilities decrease fastest for the expensive predictors.  Once all 3 of the expensive predictors' cost-adjusted posterior inclusion probabilities have decreased to 0, the cost-adjusted posterior inclusion probabilities for the non-null cheap predictors begin to decrease more quickly as $b$ increases.

Note that the choice of which prior, LCP or ECP, to use for the inclusion path should depend on the data set and overall budget. In general, for $b$ between 0 and 1 the LCP decreases prior inclusion probability and the resulting posterior inclusion probabilities for non-baseline predictors at a faster rate than the ECP does, much as a straight line is steeper than an exponential curve close to the origin. But for $b > 1$ the ECP decreases the prior inclusion probabilities more quickly, as the exponential rate surpasses the slope of the linear function.  If a degree of cost penalization less than that provided by the FND prior is desired, the ECP may produce a more useful inclusion path, as the posterior inclusion probabilities will change more gradually for $0 < b < 1$, creating a more distinct ordering for the individual inclusion paths.

\renewcommand{\thefigure}{S.\arabic{figure}}

\begin{figure}[]
\centering
\begin{subfigure}[b]{0.5\textwidth}
  \centering
  \includegraphics[width=1\linewidth]{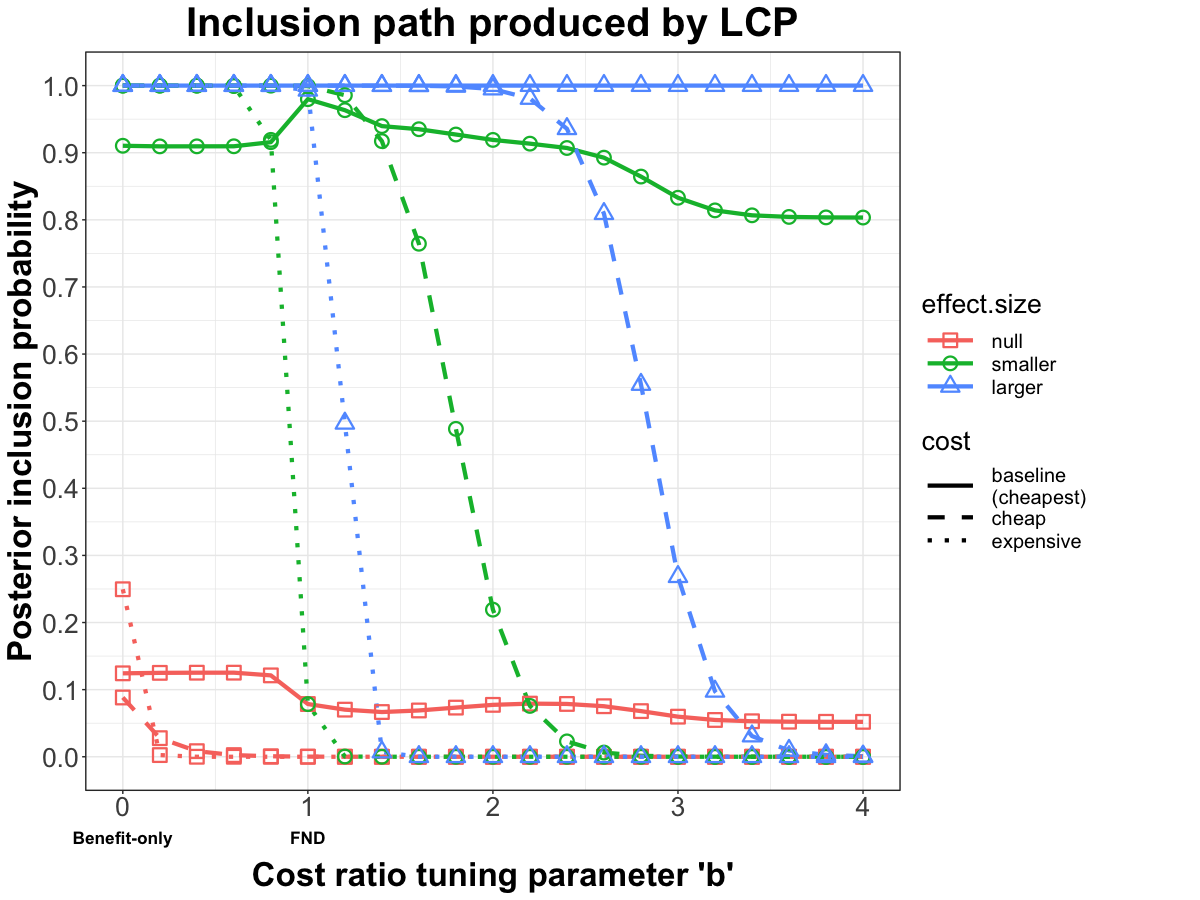}
  \caption{}
  \label{fig:inclusion}
\end{subfigure}
\begin{subfigure}[b]{0.5\textwidth}
  \centering
  \includegraphics[width=1\linewidth]{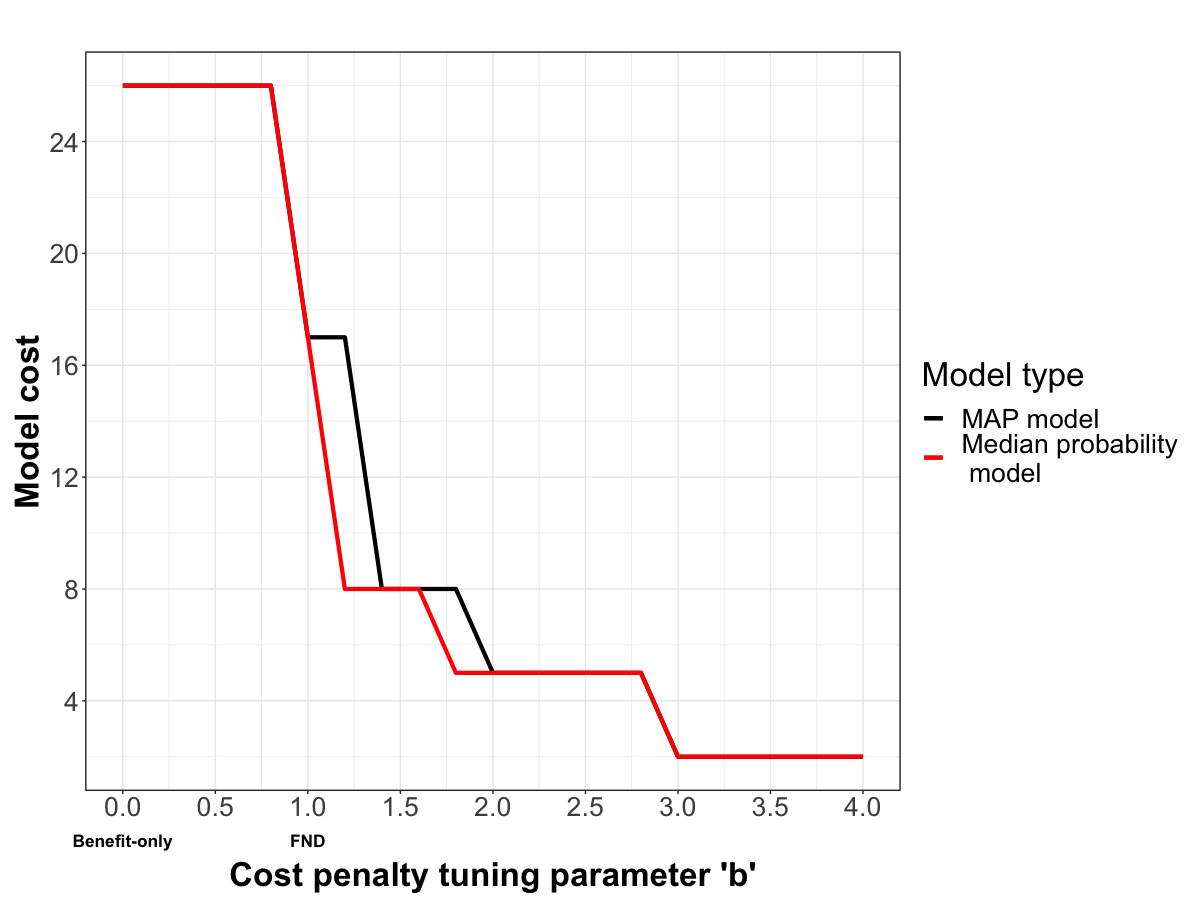}
  \caption{}
  \label{fig:cost}
\end{subfigure}
\begin{subfigure}[b]{0.5\textwidth}
  \centering
  \includegraphics[width=1\linewidth]{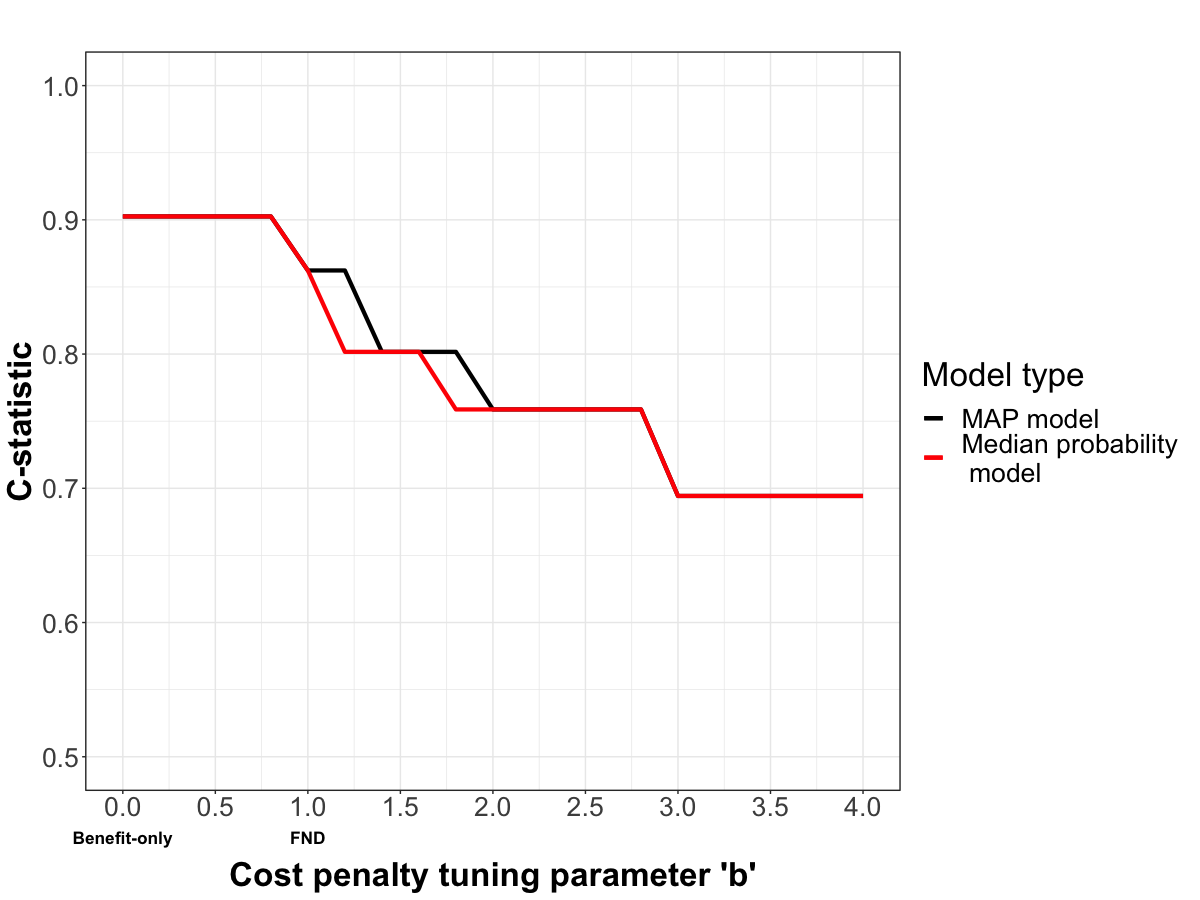}
  \caption{}
  \label{fig:cstat}
\end{subfigure}
\caption{(a) Inclusion paths for each of the $9$ predictors with baseline, cheap, and expensive costs and null, smaller, and larger effect sizes for the $n=450$ data set used in Figure 2 of the manuscript, (b) the cost of the selected model, and (c) the C-statistic of the selected model.  Model selection was performed using the LCP and tuning parameter values $b$ from 0 and 4 in increments of 0.1.  Values of $b$ above $4$ are not shown here because there is no further change in the posterior inclusion probabilities or selected models.}
\label{fig:3plots}
\end{figure}

\begin{figure}[h!]
\centering
\begin{subfigure}[b]{0.5\textwidth}
  \centering
  \includegraphics[width=1\linewidth]{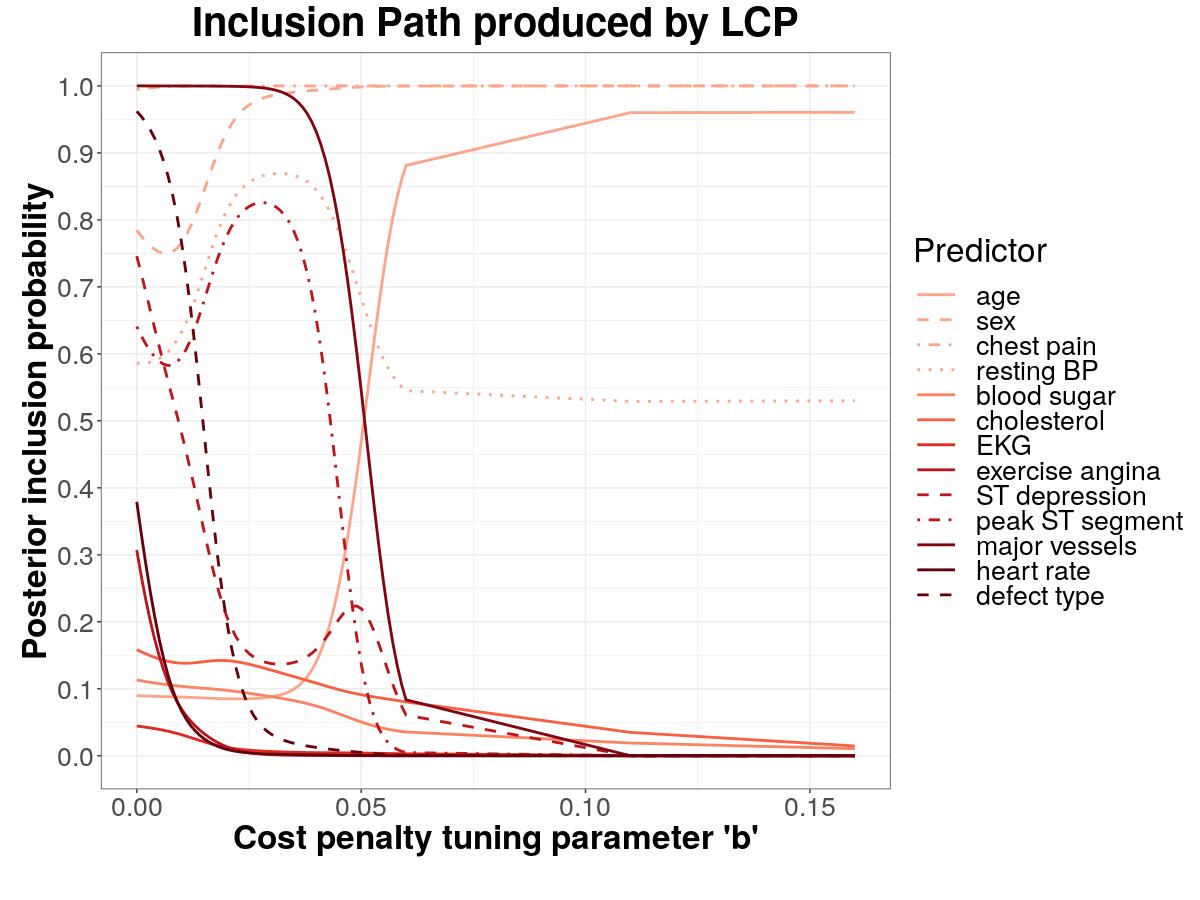}
  \caption{}
  \label{fig:heart_inclusion}
\end{subfigure}
\begin{subfigure}[b]{0.5\textwidth}
  \centering
  \includegraphics[width=1\linewidth]{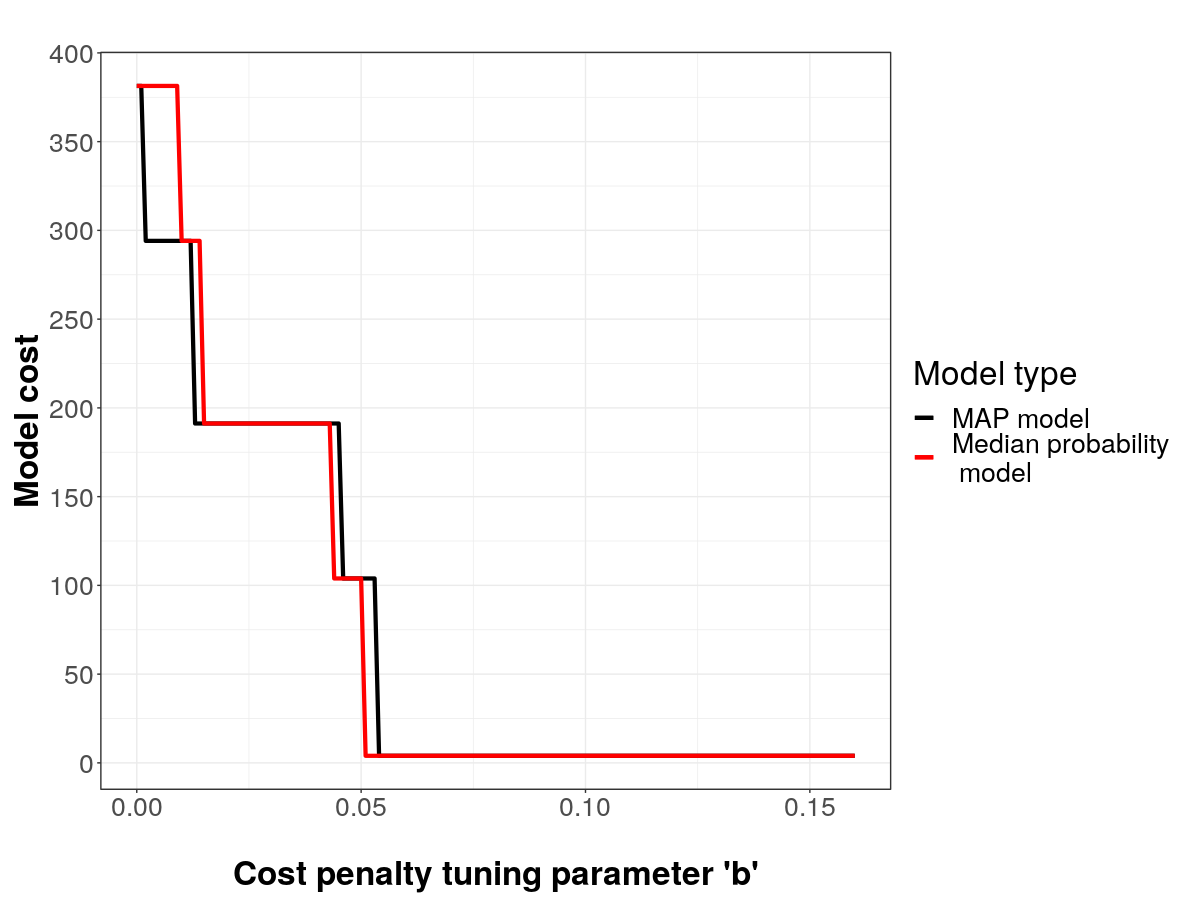}
  \caption{}
  \label{fig:heart_cost}
\end{subfigure}
\begin{subfigure}[b]{0.5\textwidth}
  \centering
  \includegraphics[width=1\linewidth]{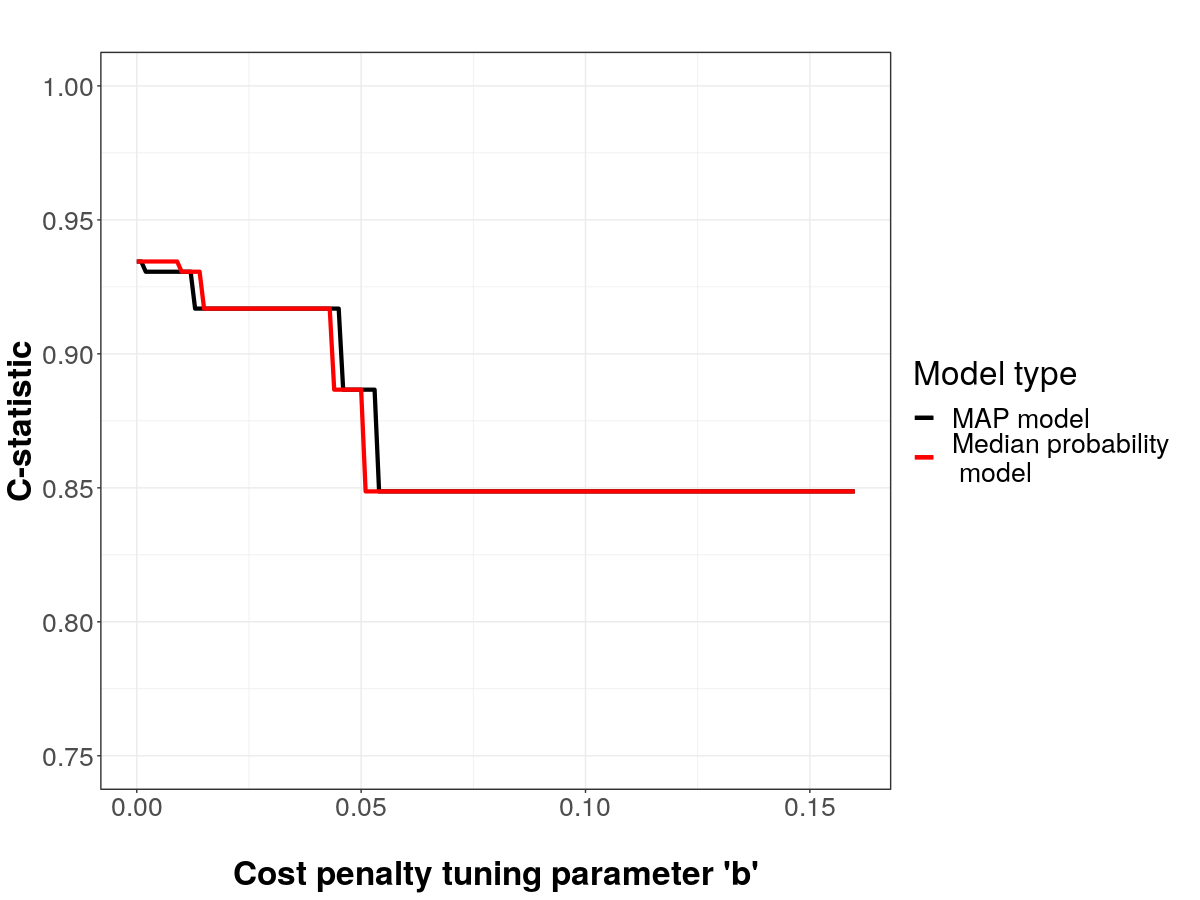}
  \caption{}
  \label{fig:heart_cstat}
\end{subfigure}
\caption{(a) Inclusion paths for each of the $13$ predictors from the heart disease data, (b) the cost of the MAP and median probability models, and (c) the C-statistic of the MAP and median probability models.  Model selection was performed using the LCP and tuning parameter values $b$ between 0 and 0.2.}
\label{fig:heart_3plots}
\end{figure}

Figure \ref{fig:heart_3plots} contains plots for the inclusion path and cost and C-statistic of the selected models using the LCP to analyze the heart disease data described in Section 3.3 of the manuscript.  The inclusion path trends in Figure \ref{fig:heart_3plots}(a) are the same as those produced by the ECP in Figure 6(a), but the cost-adjusted posterior inclusion probabilities stop changing with $b$ around $b=0.2$ since the cost penalty from the LCP is greater between 0 and 1.  

\subsection{Correlated predictors}
\begin{figure}[]
\centering
\begin{subfigure}[b]{0.5\textwidth}
  \centering
  \includegraphics[width=1\linewidth]{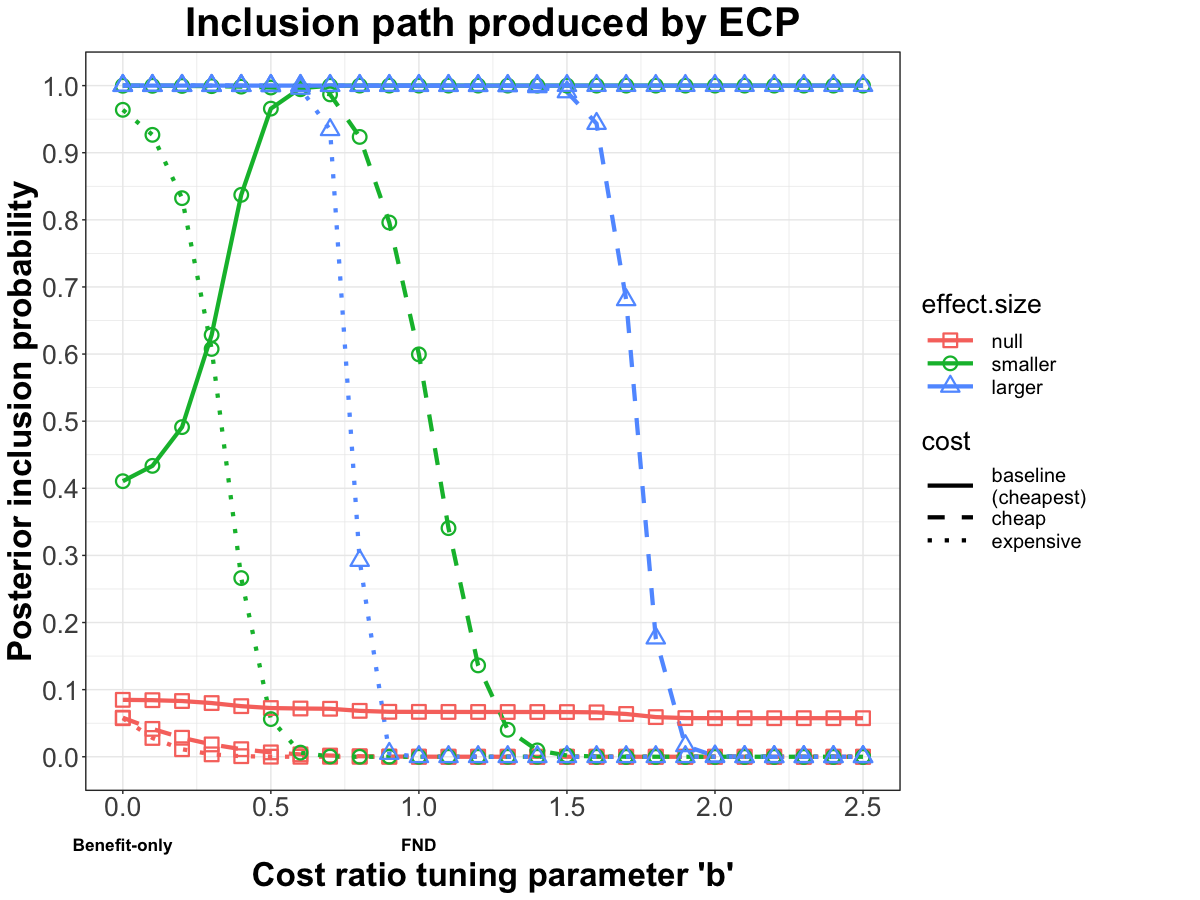}
  \caption{}
  \label{fig:inclusion}
\end{subfigure}
\begin{subfigure}[b]{0.5\textwidth}
  \centering
  \includegraphics[width=1\linewidth]{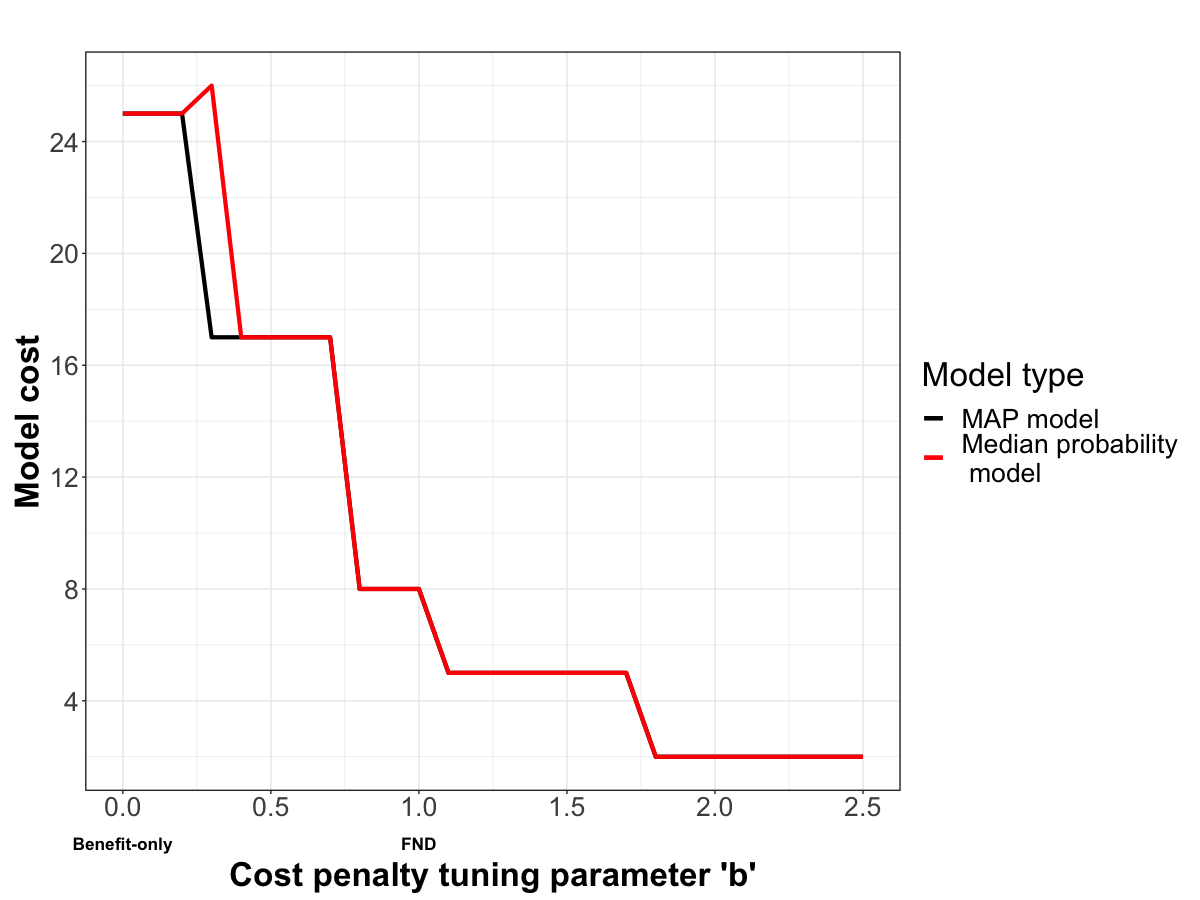}
  \caption{}
  \label{fig:cost}
\end{subfigure}
\begin{subfigure}[b]{0.5\textwidth}
  \centering
  \includegraphics[width=1\linewidth]{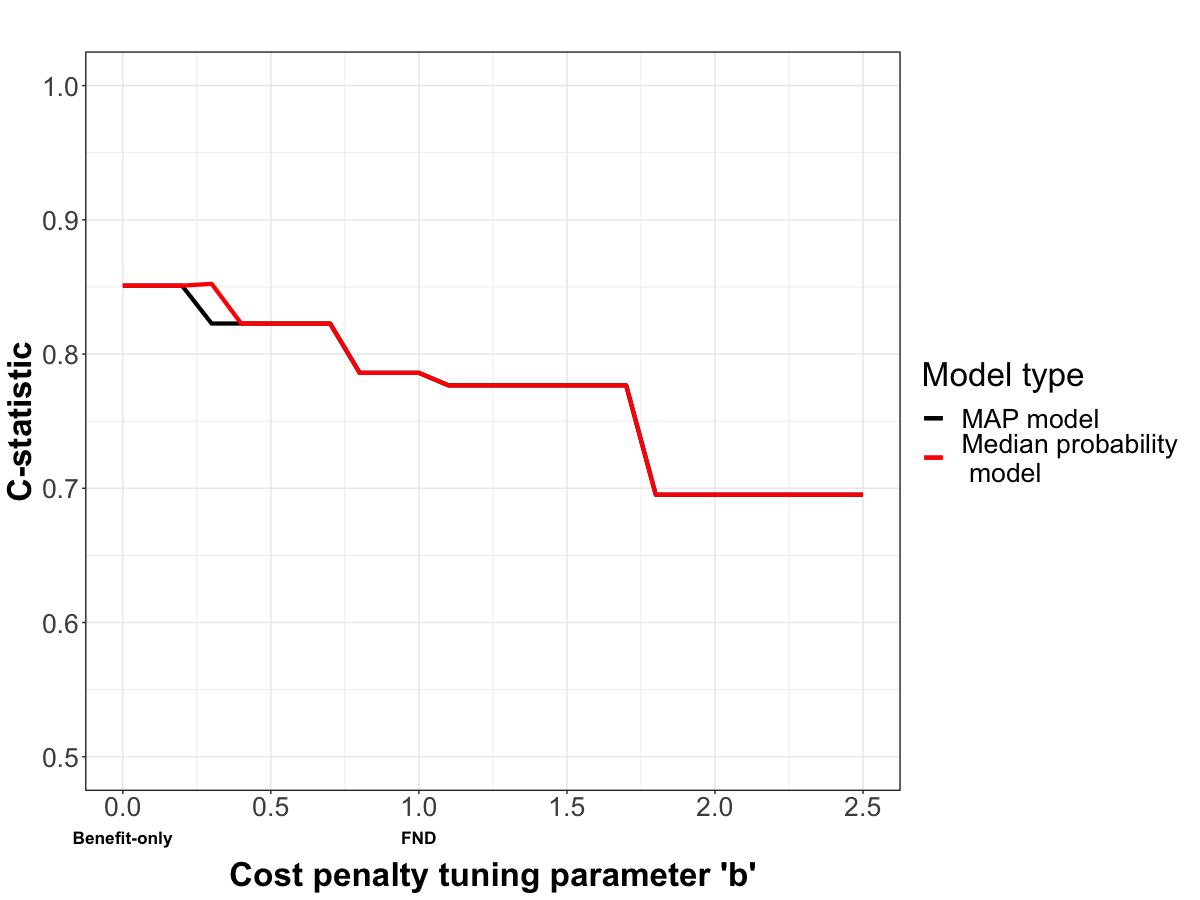}
  \caption{}
  \label{fig:cstat}
\end{subfigure}
\caption{Plots for a $n=450$ data set with $9$ predictors with baseline, cheap, and expensive costs and null, smaller, and larger effect sizes. The baseline and expensive predictors with the smaller effect size have correlation equal to 0.7. 
 (a) Inclusion paths for each of the 9 predictors, (b) the cost of the selected model, and (c) the C-statistic of the selected model.  Model selection was performed using the ECP and tuning parameter values $b$ from 0 and 2.5 in increments of 0.1.  Values of $b$ above $2.5$ are not shown here because there is no further change in the posterior inclusion probabilities or selected models.}
\label{fig:3plots_correlated}
\end{figure}

\textcolor{black}{To examine how correlated predictors might impact cost-penalized selection results, we simulated a data set using the settings outlined in Section 2.5 of the manuscript, setting the correlation to 0.7 between the baseline and expensive predictors with the smaller effect size (i.e. $X_4$ and $X_6$ as listed in Tables 1 and 2 of the manuscript).  We used a sample size $n=450$, similar to the data studied in Section 3.2.  Figure S.3 contains plots of the inclusion paths and the cost and C-statistic of the selected model for this data across different values of $b$.
Figure \ref{fig:3plots_correlated} shows that, for this particular data set, the posterior inclusion probability of the expensive predictor with smaller effect size is higher than that of the baseline predictor with smaller effect size in the benefit-only analysis.  As the cost penalty is introduced and then increases with $b$, the cost-adjusted posterior inclusion probability of the expensive predictor with smaller effect size decreases towards zero, while the cost-adjusted posterior inclusion probability of the baseline predictor with smaller effect size increases to 1.  This seems reasonable, since the baseline predictor is not penalized for its cost and thus has higher prior probability than the expensive predictor of the same effect size with whom it is correlated in the selected model.  We can see that the cost and C-statistic of the MAP model and median probability model for this data set differ briefly, specifically at $b=0.3$ when both of the correlated predictors have cost-adjusted posterior inclusion probabilities above 0.5 but the expensive predictor with smaller effect size is not included in the model with the highest cost-adjusted posterior model probability.}

